\documentclass[a4paper,twocolumn,11pt,accepted=2021-07-09]{quantumarticle}
\pdfoutput=1

\usepackage{dsfont}
\usepackage{amsmath}
\usepackage{amssymb}
\usepackage{physics}
\usepackage{float}
\usepackage{graphicx}
\usepackage{dcolumn}
\usepackage{bm}
\usepackage{hyperref}
\usepackage[backend=bibtex,giveninits=true,doi=true,isbn=false,url=false,eprint=false, style = numeric-comp, sorting=none]{biblatex}
\renewbibmacro{in:}{}
\usepackage{verbatim}
\usepackage{enumitem}
\usepackage{mathtools}
\usepackage[caption=false]{subfig}
\usepackage[utf8]{inputenc}
\usepackage[english]{babel}
\usepackage[T1]{fontenc}

\begin{document}

\title{Effect of environment on the interferometry of clocks}

\author{ Harshit Verma}
 \email{h.verma@uq.edu.au}
\affiliation{Centre for Engineered Quantum Systems, School of Mathematics and Physics, The University of Queensland, St Lucia, QLD 4072, Australia}
\orcid{0000-0002-6177-5104}

\author{Magdalena Zych}
\affiliation{Centre for Engineered Quantum Systems, School of Mathematics and Physics, The University of Queensland, St Lucia, QLD 4072, Australia}
\orcid{0000-0002-8356-7613}

\author{Fabio Costa}
\email{f.costa@uq.edu.au}
\affiliation{Centre for Engineered Quantum Systems, School of Mathematics and Physics, The University of Queensland, St Lucia, QLD 4072, Australia}
\orcid{0000-0002-6547-6005}


\begin{abstract}
Quantum interference of ``clocks'', namely of particles with time-evolving internal degrees of freedom (DOF), is a promising avenue to test genuine general relativistic effects in quantum systems. The clock acquires which path information while experiencing different proper times on traversing the arms of the interferometer, leading to a drop in its path visibility.  We consider scenarios where the clock is subject to environmental noise as it transits through the interferometer. In particular, we develop a generalized formulation of interferometric visibility affected by noise on the clock. We find that, for small noise and small proper time difference between the arms, the noise further reduces the visibility, while in more general situations it can either increase or reduce the visibility. As an example, we investigate the effect of a thermal environment constituted by a single field mode and show that the visibility drops further as the temperature is increased. Additionally, by considering noise models based on standard quantum channels, we show that interferometric visibility can increase or decrease depending on the type of noise and also the time scale and transition probabilities. The quantification of the effect of noise on the visibility -- particularly in the case of a thermal environment paves the way for a better estimate on the expected outcome in an actual experiment.
\end{abstract}
\maketitle

\section{Introduction}

One of the frontiers of modern physics has been to explore the interplay between quantum mechanics and gravity and possibly arrive at a theory of quantum gravity.
While observing this interplay in the high energy regime remains out of reach, there has been a recent focus on the low energy regime \cite{Zych2011,Zych2016,Carney_2019,Bose,Vedral,Markus,Krisnanda2020, Bose2,unknown, Guerreiro_2020}. Here, relativistic effects may be probed in quantum systems in the near future as the standards of control and metrological precision in the quantum domain have been improving swiftly and consistently~ \cite{macrosup,atom1,Fein2019,Xu745,Westphal2021}. Moreover, quantum instruments and protocols involving space based platforms are susceptible to relativistic effects and therefore, they must be appropriately accounted for in the functioning of such systems \cite{sat1,sat2,sat3} which are slated to be operational in the near future.

The experimental tests probing the effect of gravity on quantum systems which have been performed so far, such as neutron interferometry \cite{neutro1,neutro2}, are only sensitive to the non-relativistic, Newtonian potential. As a probe of general relativistic effects, Refs.~\cite{Zych2011,Zych2016} proposed interference of particles with dynamically evolving internal DOF (such as constituted by the vibrational levels of a molecule or the energy levels of an atom). Thinking of the internal DOF as ``clocks'' tracking the particle's proper time, one can effectively probe a ``superposition of proper times'' by placing the particle in a spatial superposition. In the proposed experiment, one such particle travels in a state of superposition through the arms of a Mach-Zehnder (MZ) interferometer. The state of the clock residing inside the particle evolves through proper time which is dependent on the path taken by the centre of mass of the particle, effectively leading to entanglement of the clock with the centre of mass of the particle. There exists a complementarity relation between path distinguishability ($\mathcal{D}$) and interferometric visibility ($\mathcal{V}$) \cite{complement,Zhou_2018} which is given as follows: 
\begin{equation}
\mathcal{V}^2 + \mathcal{D}^2 \leq 1~.
\label{eq:inequality}
\end{equation}
Therefore, the acquisition of which-path information by the clock increases the path distinguishability and hence, leads to a drop in the interferometric visibility measured on the path.

Soon enough, a number of experimental realizations of the above proposal were laid out based on different physical systems \cite{singleE,singleP,Zych_2012,atom1,atom2,2019arXiv,unstable}. In the spirit of the original proposal, an important experiment using cold atom condensates was done in \cite{Marg1205,Zhou_2018} which provided a proof of principle by simulating the difference in ticking rates of clocks using external magnetic fields. The effect was also proposed as a possible universal decoherence mechanism due to gravitational interaction \cite{pikovski2015, pikovski2017}. 

In general, all the clock interferometry experiments proposed so far assume that there is no noise acting on any of the DOF. But it can be realistically expected that the DOF associated with the particle are susceptible to environmental effects such as noise.

There are two qualitatively distinct types of noise one can consider: on the path and on the internal degrees of freedom. Noise acting on the path is a common feature of all interference experiments and results in the well-known loss of visibility due to the direct dephasing of the spatial superposition \cite{dephase}. Noise acting on the internal degrees of freedom is more interesting. In the absence of time dilation, this type of noise would have no effect on the path visibility. This is because, as long as no other coupling exists between path and internal DOF, the two remain decoupled, so that noise on the second has no effect on the first. This fact was attested experimentally in Ref. \cite{Fein2019} using molecules with different properties which led to no change in their interference pattern. Therefore, any effect of internal noise on path visibility must take the relativistic aspect into account.

Furthermore, one might develop contrasting intuitions on the effect of noise acting on internal DOF. On one hand, we may be inclined to think that this will lead to an increase in the visibility as the which-path information acquired by the clock gets corrupted by the noise. On the other hand, we may infer that it may lead to a decrease in visibility because the environment can be modelled as additional DOF to the clock, thus, increasing the dimensionality of the clock system (clock+environment). As a result, it may be expected that for a given proper time difference, the clock + environment system can better note the which-path information, thus leading to a higher distinguishability, and hence a lower visibility. All of the above inferences have been drawn using the complementarity relation in Eq.~\eqref{eq:inequality}.

In the light of the above contradictory intuitions and the experimental relevance, it is important to understand if and how noise acting on internal DOF can affect the visibility. Here we develop a general model for studying how the interferometric visibility is affected by noise acting on the clock. We apply our method to various external environments acting on the clock and find that depending on different parameters, the visibility can decrease or increase. We first consider the clock to be interacting with a single mode of bosonic bath which is modelled as Jaynes-Cummings (JC) model \cite{JCmodel}. As an interesting case, we also consider a finite temperature environment in the JC model, which is similar to the experimental condition of a clock in a cavity at non-zero temperature. Next, we use standard quantum channels -- amplitude damping (AD), phase damping (PD) and depolarizing (DP) \cite{chuang,open} to model the noisy evolution of the clock. We find that in the low noise regime, the visibility drops universally (in all the models) for a small proper time difference. This can be intuitively expected as the effect of the extended Hilbert space of the clock system overwhelms the loss in which path information.Our formalism also allows for the case wherein different environments (though qualitatively similar), act on each of the two arms of the interferometer. In this case, the environment can acquire which-path information, resulting in a visibility loss that does not depend uniquely on relativistic effects. Nonetheless, as it can be interesting to explore the interplay between the two effects. Therefore, we explore this case in channel based noise models.

\section{Theoretical Framework}
\begin{figure}
    \includegraphics[trim={5.5cm 9.25cm 4cm 12.5cm},clip,width=1\columnwidth]{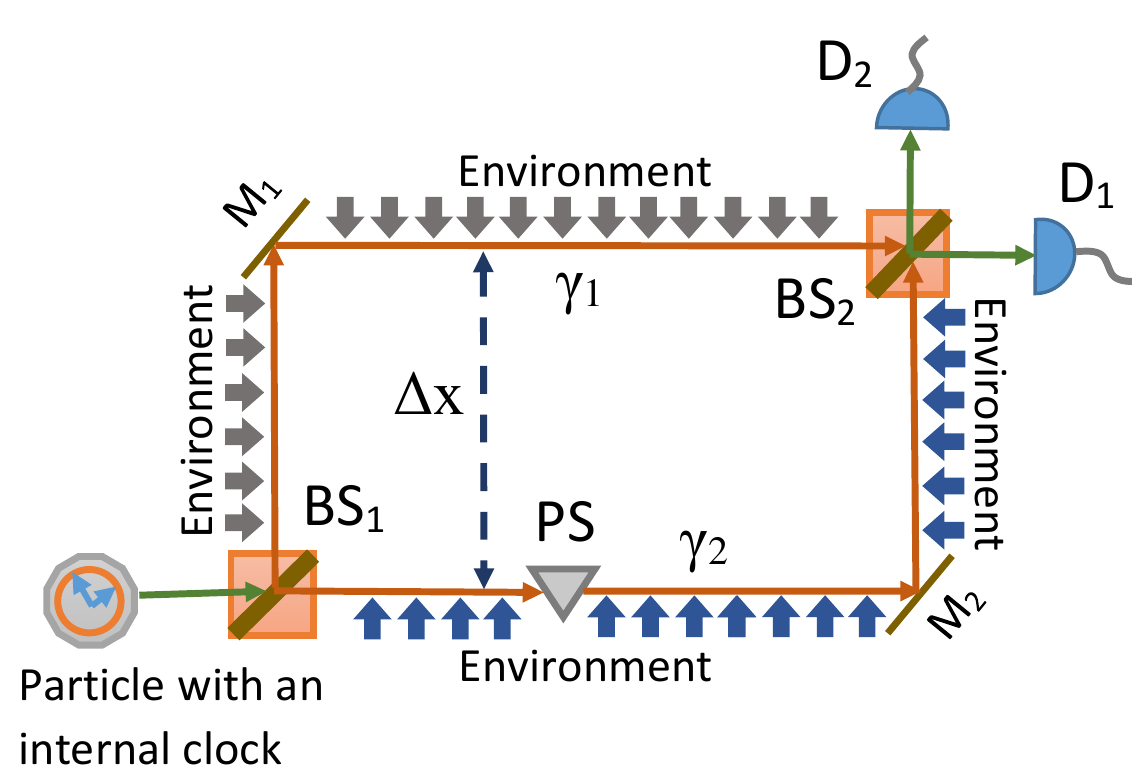}
    \caption{Schematic diagram of the proposal with Mach-Zehnder interferometer. The massive particle with an internal clock is to the left. The beamsplitters are denoted by $BS_1$ and $BS_2$ and the phase-shifter is denoted by $PS$. The spatial separation in the semi-classical trajectory of centre of mass of the particle through the arms of the interferometer is shown as $\Delta x$. Detectors measuring the port from which the particle emerges are denoted by $D_1$ and $D_2$ respectively. $M_1$ and $M_2$ are mirrors which have no effect on the evolution of the internal DOF. The different colors denoting environmental effects through both the paths ($\gamma_1$ and $\gamma_2$) indicate that the arms are not coupled to the same environment but rather to similar but independent environments.}
    \label{fig:MZ}
\end{figure}
A simple setup to study relativistic effects on quantum interferometric visibility is a MZ type interferometer as shown in Fig.~(\ref{fig:MZ}). In this setup, a massive particle is prepared in a spatial superposition as it goes through the first beamsplitter -- $BS_1$. Such a particle is assumed to contain an internal DOF whose dynamics is governed by the corresponding internal Hamiltonian $H_{\textrm{0}}$. This internal DOF is initialized in a state $\ket{\psi_{\textrm{clock}}}$, which is not an eigenstate of $H_\textrm{0}$. This condition ensures that the internal DOF have a non-trivial time evolution and therefore effectively function as a ``clock'', in the sense that they go through distinguishable states with the passing of time \footnote{The use of the word ``clock'' is meant to be suggestive of the system's sensitivity to proper time, in reference to the operational understanding of time in relativity, with no further implications. Even though any time-evolving system can be used in principle to measure time, we are not interested here in constructing practical clocks, but simply in joint signatures of quantum and relativistic effects. In fact, our main quantity of interest is visibility of the path DOF, which does not require reading off the clock's time.}. As it traverses the arms of the MZ interferometer, the clock experiences different proper times. Therefore, if the internal DOF is prepared in a state $\ket{\psi_{\textrm{clock}}}$ before $BS_1$, it will be evolving as follows:
\begin{align}
\ket{\tau} := e^{-iH_{\textrm{0}} \tau}\ket{\psi_{\textrm{clock}}} , 
\end{align} 
along an arm with the proper time $\tau$. It is to be noted that $\hbar=1$ has been assumed throughout the manuscript. On the paths $\gamma_1$ and $\gamma_2$, the state of the particle is denoted by $|1\rangle$ and $|2\rangle$ respectively, and $BS_1$ prepares a state $\frac{|1\rangle + |2\rangle}{\sqrt{2}}$. Including the clock leads to the following entangled state of the system after time evolution, just before it passes through the measurement apparatus consisting of the beamsplitter $BS_2$ and particle detectors $D_1$ and $D_2$:
\begin{align} \label{noiseless}
|\psi_\textrm{out}\rangle = \frac{1}{\sqrt{2}} [\ket{1}\ket{\tau_1} +e^{i \chi}\ket{2}\ket{\tau_2}] , 
\end{align}
where $\tau_1$ and $\tau_2$ are the proper times along the paths $\gamma_1$ and $\gamma_2$ respectively \cite{Zych2011}. One of the arms of the interferometer ($\gamma_2$) also has a phase-shifter ($PS$) which introduces a controllable phase of $\chi$. This can be used for the measurement of the phase difference between the states of the particle in the two arms of the interferometer. During the transit of the particle through the arms of the interferometer, the environment is assumed to act on the internal DOF of the particle i.e.\ the clock. Thereafter, the particle goes through the measurement apparatus consisting of the beamsplitter $BS_2$ and the detectors $D_1$ and $D_2$. Over many experimental runs, the number of particles detected by $D_1$ and $D_2$ gives an estimate of the probabilities associated with each path which are further used to calculate the visibility through the following equation:
 \begin{align}
\mathcal{V} = \frac{Max_\chi P_{\pm} - Min_\chi P_{\pm}}{Max_\chi P_{\pm}+Min_\chi P_{\pm}}~,
\label{eq:visi}
\end{align}
where $P_+$ and $P_-$ are probabilities of detection through detectors $D_1$ and $D_2$ respectively.

For the sake of completeness, we highlight the previous results for the following two cases (assuming no noise acting on the path):
\begin{itemize}[leftmargin=0.25 cm]
\item \textbf{No notion of proper time, i.e., no internal/evolving DOF}:
$\mathcal{V} = 1$. This case is simply the gravitational analogue of the electric Aharonov-Bohm effect \cite{AB}.
\item \textbf{With the notion of proper time but without the environment acting on the clock}: $\mathcal{V}=|\langle \tau_1|\tau_2\rangle|$ \cite{Zych2011}. Here, $|\tau_1\rangle$ and $|\tau_2\rangle$ are the states of the internal DOF after going through the paths $\gamma_1$ and $\gamma_2$ (which refer to the two arms of the interferometer as shown in Fig.~\ref{fig:MZ}) and consequently evolving with the proper times $\tau_1$ and $\tau_2$ respectively. The visibility can also be expressed in terms of the proper time difference ($\Delta \tau = \tau_2-\tau_1$) between the paths by the following relation \cite{pikovski2017}: 
\begin{equation}
\mathcal{V}=|\bra{\psi_{\textrm{clock}}} e^{-iH_{\textrm{0}} \Delta\tau}\ket{\psi_{\textrm{clock}}}|.
\label{eq:diff}
\end{equation}
\end{itemize}

In general, it may be assumed that the arms of the interferometer are two distinct worldlines through space-time having different proper times along them. However, the time of arrival of the wave packets associated with the particle along the distinct paths at $BS_2$ should  be the same. This peculiarity can be understood as arising from different frames of reference -- one is the fixed laboratory frame and the second is the frame co-moving with the particle's amplitude which has proper time along the path as its time coordinate. 

A specific example of the proposal is shown in Fig.~\ref{fig:MZ} wherein the centre of mass of the massive particle is assumed to follow a semi-classical trajectory through the arms of the interferometer which are separated by a spatial distance $\Delta x$ and are kept in a homogeneous gravitational field perpendicular to the arms of the interferometer. In the field, the separation ($\Delta x$) would give rise to different relativistic effects in the arms of the interferometer -- particularly different time dilation and hence, different proper times. In such a case, the proper time difference would be $\Delta \tau =  \frac{g}{c^2}\Delta x~t$, where $g$ is the acceleration due to gravity and $t$ the time in laboratory frame. 
\section{Visibility due to noisy clocks}
In this section, we assess the effect of the environment on the visibility in a generic scenario wherein noise acts on the clock. We reiterate that we do not consider noise which only acts on the path DOF of the particle in which case the visibility simply decreases as the noise is increased.

To describe the dynamics of noisy clocks, one of the approaches could be that of solving the master equation for the clock corresponding to the noise. However, it is not a priory clear that we can  apply the master equation approach in a scenario where different amplitudes of a particle (here, corresponding to the amplitudes in spatial superposition of the particle containing clock) evolve with different proper times, as it is in the case of the two arms of a MZ interferometer \cite{pikovski2015}. Noting that the final effect depends on the total proper times (their difference) on each of the paths, the proposition of Markovian evolution -- essential for the master equation approach, looks untenable. Moreover, the quantum interference after evolution through the arms of the MZ further complicates the implementation of Master equation approach.

Therefore, we resort to the basic theory of open quantum systems where the noise is understood as arising from the target-environment interactions. The extended system (target + environment) undergoes a unitary evolution as a whole and the non-unitary, noisy dynamics of the target can be obtained by tracing out the environment \cite{open}. In general, the environment could be constituted by quantum DOF which could be bosonic or fermionic.

In light of the above arguments, we consider below a generic case to obtain an expression for visibility in the presence of an environment, i.e. with the noisy dynamics of the clock. For simplicity, we have assumed that the clocks in our case are  initially prepared in a pure state. The extension to multi-level clocks or mixed states is straightforward \cite{pikovski2015}. Initially, the clock is prepared in the superposed state of its energy eigenstates so that their evolution can be used for keeping  track of proper time. 

The clock is susceptible to a variety of physical processes which are modelled as noise, e.g., a two level clock in its excited state may decay to the ground state or the coherence in the superposition of the clock state may be lost due to dephasing. In the presence of noise, the dynamics of the clock is governed by its intrinsic Hamiltonian (henceforth denoted by $H_\textrm{0}$) as well as the noise Hamiltonian ($H_\textrm{noise}$). $H_\textrm{noise}$ represents the interaction of the target i.e.\ clock with the environment capturing the possible effects of noise. In general, the environmental DOF could also be interacting among themselves or with other external DOFs, the corresponding dynamics being governed by $H_\textrm{env}$. Therefore, the total Hamiltonian governing the dynamics of the clock is $H_\textrm{int}=H_0 + H_\textrm{noise}$. If we include the environment with the clock to consider the evolution of the extended system (clock + environment) to be governed by a unitary time evolution operator $U_\textrm{int}$, the corresponding Hamiltonian governing its dynamics would be: $H_\textrm{int}=H_0 + H_\textrm{noise} + H_\textrm{env}$. 

\subsection{Expression for visibility in presence of a generic environment and interaction}
To derive the expression for the visibility in the presence of an environment, as a simple case, we can consider an environment which is initialized in a pure state $|\alpha\rangle_E$ along with the clock initialized in the state $|\psi\rangle_0$. A case where the clock is initially in a mixed state will be considered in the next section. Therefore, the initial state of the combined system is given as follows:
\begin{align}
|\psi\rangle_\textrm{CE}(t=0) = |\psi_0\rangle_C \otimes |\alpha\rangle_E~.
\label{eq:init}
\end{align}
Henceforth, we suppress the subscripts ``C" for clock DOF, and ``E" for the environment DOF, as well as the tensor product symbol. Therefore, Eq.~\eqref{eq:init} becomes $|\psi\rangle = |\psi\rangle_0|\alpha\rangle$. Adopting the above convention, in the subsequent equations the first ket will represent the clock and the second will represent the environment.

As discussed above, we are interested in how noise acting on the clock alone affects the visibility in the path DOFs. Therefore, we assume that $H_{\textrm{noise}}$ acts non-trivially only on the environment and the clock, while it acts as identity on the path. The clock + environment DOF in both the arms experience a unitary evolution governed by the time evolution operator $U_\textrm{int} = e^{-iH_\textrm{int}t}$ and if we assume that proper time through $\gamma_1$ is $\tau_1$ and that through $\gamma_2$ is $\tau_2$, the generic state of this extended system at later times is given as follows:
\begin{align}
 |\psi_{\gamma_1}(\tau_1)\rangle&= e^{-iH_\textrm{int}\tau_1} \left(|\psi\rangle_0|\alpha\rangle\right) \nonumber\\
|\psi_{\gamma_2}(\tau_2)\rangle &= e^{-iH_\textrm{int}\tau_2} (|\psi\rangle_0|\alpha\rangle)~,
\label{coeff}
\end{align}
which, in general, can be an entangled state of the clock and environment DOF, given that $H_\textrm{noise}\neq 0$.

Considering the evolution of the system through various parts of the interferometer including the beamsplitter $BS_1$, phaseshifter $PS$ and the mirrors, we find the following state just before the particle goes through the detection system consisting of $BS_2$ and detectors $D_1$ and $D_2$ (Fig.~\ref{fig:MZ}):
\begin{align}
|\psi\rangle = \frac{1}{\sqrt{2}} \big[&i e^{-i\phi_1}|1\rangle_\gamma |\psi_{\gamma_1}(\tau_1)\rangle \nonumber\\
&+ e^{-i\phi_2 +i\chi}|2\rangle_\gamma|\psi_{\gamma_2}(\tau_2)\rangle\big],
\label{eq:state}
\end{align}
where $\phi_1$ and $\phi_2$ correspond to Aharonov-Bohm type phases introduced by virtue of the different gravitational potentials along each of the paths and $\Delta\phi=\phi_1-\phi_2$. In terms of the path DOF, $|1\rangle_\gamma \leftrightarrow \gamma_1$ and $|2\rangle_\gamma \leftrightarrow \gamma_2$. After constructing the density matrix of the state in Eq.~\eqref{eq:state}, ($\rho=|\psi\rangle\langle\psi|$) and tracing out the clock and environmental DOF (so that only the path DOF of the particle may be considered), we obtain,
\begin{align}
\rho_\textrm{out} &= \trace_\textrm{C,E}[\rho] \nonumber\\
&= \frac{1}{2} \big[ |1\rangle_\gamma\langle 1|_\gamma \langle\psi_{\gamma_1}(\tau_1)|\psi_{\gamma_1}(\tau_1)\rangle \nonumber \\
&+ |2\rangle_\gamma \langle2|_\gamma \langle\psi_{\gamma_2}(\tau_2)|\psi_{\gamma_2}(\tau_2)\rangle\nonumber\\
&-i e^{i(\Delta \phi + \chi)}|2\rangle_\gamma\langle1|_\gamma \langle\psi_{\gamma_1}(\tau_1)|\psi_{\gamma_2}(\tau_2)\rangle \nonumber \\
&+i e^{-i(\Delta \phi +\chi)}|1\rangle_\gamma\langle2|_\gamma\langle\psi_{\gamma_2}(\tau_2)|\psi_{\gamma_1}(\tau_1)\rangle\big]. \nonumber
\end{align}
We obtain the probability of getting a signal on either of the detectors, $P_{\pm}$ as follows:
\begin{align}
P_{\pm}&=\langle\pm|\rho_\textrm{out}|\pm\rangle \nonumber\\
&= \frac{1}{2}\bigg[1 \pm \frac{ |\kappa|}{2} \big( e^{i(\Delta \phi + \chi+\Upsilon-\pi/2)}\nonumber\\
&~~~~~~~~~+ e^{-i(\Delta \phi + \chi+\Upsilon-\pi/2)} \big)  \bigg] \nonumber\\
&\equiv \frac{1}{2}\bigg[1\pm |\kappa| \sin({\Delta \phi + \chi+\Upsilon})\bigg], 
\end{align}
where $\kappa=\langle\psi_{\gamma_1}(\tau_1)|\psi_{\gamma_2}(\tau_2)\rangle=|\kappa|e^{i\Upsilon}$ i.e. $\Upsilon = \arg({\kappa})$.

Finally, using Eq.~\eqref{eq:visi}, we get the visibility as $\mathcal{V} = |\kappa|$. This result has the same generic form as the original one for the clock only, that is  $\mathcal{V} = |\langle\tau_1|\tau_2\rangle|$.  However, the states $|\tau_i\rangle$ are now described by the combined clock and environment DOF and their joint evolution for proper time $\tau_i$ is given by the operator $U_\textrm{int}$. 

Therefore, we find that even if the extended system evolves to a generic state in the extended Hilbert space, the form of visibility remains the same. This proposition holds for an environment of arbitrary dimension and for an arbitrary clock-environment interaction.

\subsection{Low noise regime}
Given the limitations of the current technology, only very small proper time differences can be achieved. For reference, the largest spatial separations achieved to date in matter-wave interferometers are still below the meter \cite{Asenbaum2017}, while the longest coherence times (achieved in different experiments) are around 20 $\mathrm{s}$. Combining these figures, and assuming an optimal geometry that maximises gravitational time dilation, results in a proper time difference $\Delta \tau = \frac{g}{c^2}\Delta x~t \approx 2 \times 10^{-15}\; \mathrm{s}$.
Therefore, the particular regime of a small difference in proper time ($\Delta \tau$) is of our particular interest owing to its significance in experiments. To probe the effects of noise in this regime, we resort to analytical analysis which does not require the exact nature of the noise or the model of environment and its interaction with the clock. Here, we make the assumption that the environment does not have an evolution of its own i.e. $H_\textrm{env}=0$. Therefore, the Hamiltonian governing the time evolution of complete system is $H_\textrm{int} = H_0 + H_\textrm{noise}$.

In the low noise limit, $\Delta \tau \left\| H_\textrm{noise}\right\| \ll 1$ and we have in Eq.~\eqref{eq:diff} (with $H_\textrm{int}$), $\langle e^{-i\Delta \tau (H_0 + H_\textrm{noise})}\rangle \approx \langle e^{-i\Delta \tau H_0}\rangle \langle e^{-i\Delta \tau H_\textrm{noise}}\rangle$ for an initial product state of the clock and environment. As $|\langle e^{-i\Delta \tau H_\textrm{noise}}\rangle|<1$, this implies $\mathcal{V}<\mathcal{V}_0$ where $\mathcal{V}_0$ is the visibility without any effect of the noise. This means that the introduction of a small amount of noise reduces the visibility. Additionally, by approximating  $\langle e^{-i\Delta \tau H_\textrm{noise}}\rangle\approx 1 - i\Delta \tau \langle H_\textrm{noise} \rangle$ we may infer that the visibility monotonically decreases with increasing noise, in the low noise regime.

\section{Clocks interacting with a single bosonic field mode}
Here, we consider the Jaynes-Cummings (JC) model representing the interaction of a single qubit (clock) with a bosonic field. As the simplest case, we can consider a single field mode to constitute the environment for which it is possible to calculate the visibility analytically, which we show below. This case qualitatively represents a physical setting where the decay of the clock in an excited state causes the excitation of a bosonic field mode. For example, this would be the case for the electronic energy levels of an atom coupled to an optical cavity. If the clock interacts with a single mode of a bosonic field, in the rotating wave approximation, the Hamiltonian of the complete system is given as follows \cite{JCmodel}:
\begin{align}
H_\textrm{int} = \underbracket{\frac{(\Delta E)\sigma_C^z}{2}}_{H_0} + \underbracket{\omega a^{\dagger}a}_{H_\textrm{env}} + \underbracket{\frac{\lambda}{2}\big(a\sigma_C^+ +a^{\dagger}\sigma_C^-\big)}_{H_\textrm{noise}}, \nonumber
\end{align}
where $\sigma_C$ are the Pauli operators corresponding to the clock and $a$ and $a^{\dagger}$ are the annihilation and creation operators of the bosonic field. The frequencies of the clock and the field are $\Delta E$ and $\omega$ respectively, and the coupling constant between the field and clock is $\lambda$.

\subsection{Bosonic field in a pure fock state $|0\rangle$}
Firstly, we take the initial state of the extended system to be a product state, as in the generic case discussed earlier: $|\psi\rangle_\textrm{CE}(t=0) = |\psi_0\rangle \otimes |0\rangle_E$ (as in Eq.~\eqref{eq:init}, with $|\psi_0\rangle  = \frac{|0\rangle+|1\rangle}{\sqrt{2}}$), where $|0\rangle_E$ denotes the ground state of the field mode, $a|0\rangle_E=0$. Denoting by $|1\rangle_E := a^{\dagger}|0\rangle_E$ the 1-particle excitation of the field, the eigenvectors in the $|\textrm{clock, field}\rangle$ basis relevant for the time evolution  are as follows \cite{JCmodel}:
\begin{align}
|v_1\rangle &= |00\rangle,\nonumber\\
|v_2\rangle &= \cos{\bigg(\frac{\alpha_0}{2}\bigg)}|10\rangle + \sin{\bigg(\frac{\alpha_0}{2}\bigg)}|01\rangle,\nonumber\\
|v_3\rangle &= \sin{\bigg(\frac{\alpha_0}{2}\bigg)}|10\rangle - \cos{\bigg(\frac{\alpha_0}{2}\bigg)}|01\rangle,\nonumber
\end{align}
\\ 
where, $\alpha_0 =\tan^{-1}\big({\frac{\lambda}{\delta}}\big)$, $\delta = \Delta E-\omega$ is the detuning frequency. Therefore, the initial state of the extended system can be expressed in terms of the above eigenvectors as follows:
\begin{align}
|\psi\rangle_\textrm{CE}(0) =\frac{1}{\sqrt{2}} \bigg[ |v_1\rangle &+  \cos{\bigg(\frac{\alpha_0}{2}\bigg)}|v_2\rangle \nonumber\\
&+ \sin{\bigg(\frac{\alpha_0}{2}\bigg)}|v_3\rangle\bigg]~. \nonumber
\end{align}
The eigenvalues corresponding to the eigenvectors $|v_1\rangle, |v_2\rangle, |v_3\rangle$ are $\frac{-\Delta E}{2}$, $\frac{\omega+\lambda_0}{2}$ and $\frac{\omega-\lambda_0}{2}$ respectively, where $\lambda_0 = \sqrt{\delta^2+\lambda^2}$. Using these, we can write the time evolved state of the extended system as follows:
\begin{align}
|\psi\rangle_\textrm{CE}(t) &= \frac{e^{i\frac{\Delta E}{2}t}}{\sqrt{2}}|v_1\rangle + \frac{e^{-i\frac{\omega+\lambda_0}{2}t}}{\sqrt{2}} \cos{\bigg(\frac{\alpha_0}{2}\bigg)}|v_2\rangle \nonumber\\
&+ \frac{e^{-i\frac{\omega-\lambda_0}{2}t}}{\sqrt{2}} \sin{\bigg(\frac{\alpha_0}{2}\bigg)}|v_3\rangle .
\label{eq:JCstate}
\end{align}
Therefore, we can obtain the visibility by getting the state in Eq.~\eqref{eq:JCstate} to evolve through different proper times and taking their scalar product:

\begin{widetext}
\begin{align}
\mathcal{V} = \frac{1}{2}\bigg|e^{i\frac{\Delta E}{2}\Delta\tau} + e^{-i\frac{\omega+\lambda_0}{2}\Delta\tau} \cos^2{\bigg(\frac{\alpha_0}{2}\bigg)} +e^{-i\frac{\omega-\lambda_0}{2}\Delta\tau} \sin^2{\bigg(\frac{\alpha_0}{2}\bigg)} \bigg|~.
\label{eq:visiJCM}
\end{align}
\end{widetext}
\begin{figure*}[!ht]
     \subfloat[The variation of $\mathcal{V}$ with $\Delta E$ and $\lambda$ showing the interplay between $H_0$ and $H_\textrm{noise}$. The visibility has been calculated with a fixed $\Delta \tau = 1 $ and $\omega=1.1$.\label{fig:JCM}]
     {
       \includegraphics[width=0.35\textwidth]{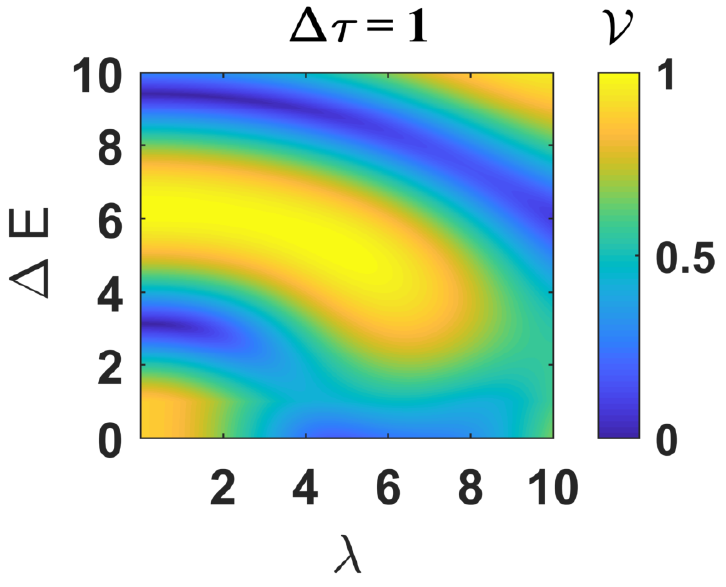}
     }
     \hfill
     \subfloat[The variation of $\mathcal{V}$ with $\lambda$ for different values of $\omega$. The visibility has been calculated with a fixed $\Delta \tau = 1 $ and $\Delta E=1$. We see that in an off-resonance condition i.e. $\Delta E \neq \omega$, the visibility is lower than in resonance condition at $\lambda=0$.\label{fig:JCMomgk}]{%
       \includegraphics[width=0.6\textwidth]{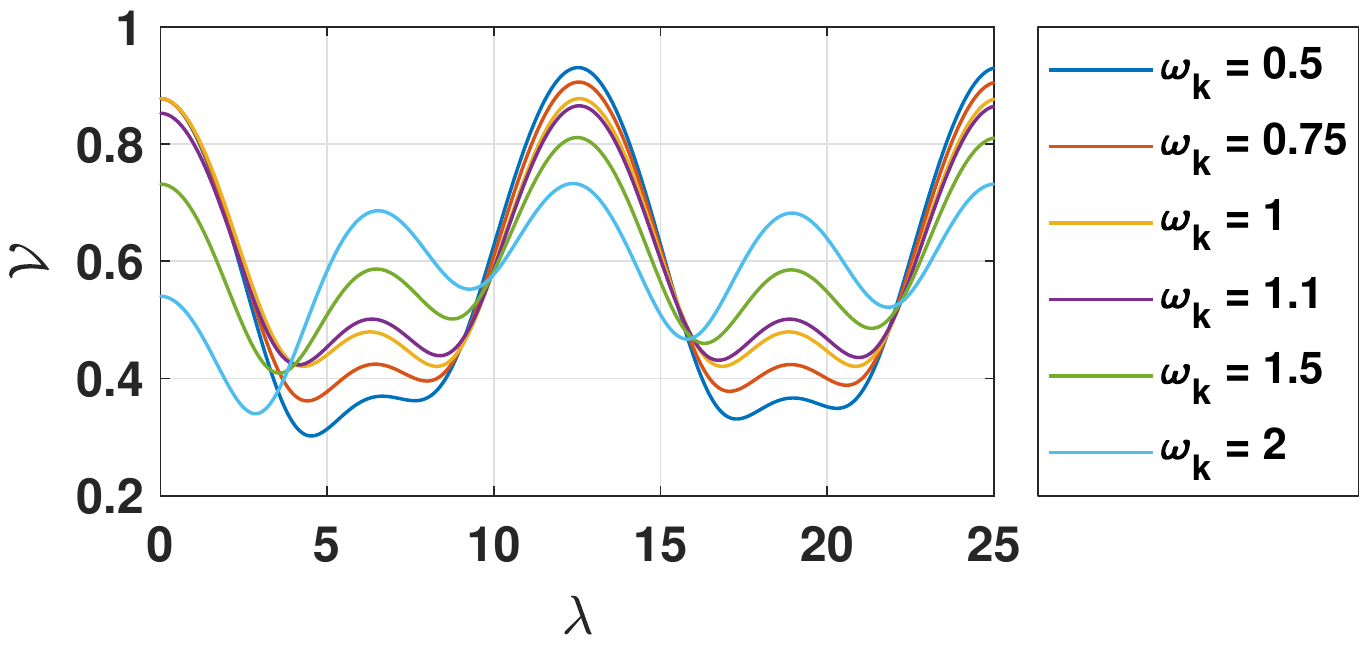}
     }
     \caption{Analysis of visibility for JC modelled environment interaction with effective temperture $T = 0$.}
\end{figure*}

Using the expression for the visibility obtained above, we show the interplay between the intrinsic dynamics of the clock, governed by $H_0$, characterized here by $\Delta E$ and the interaction between the clock and environment, governed by $H_\textrm{noise}$ and characterized here by $\lambda$ in Fig.~\ref{fig:JCM}. The interplay between aforementioned parameters leads to a change in visibility which now attains both the minimum value of zero and the maximum value of one in the different regimes of $\Delta E$ and $\lambda$. Given the dynamic nature of the bath constituted by a bosonic field of frequency $\omega$, it is also interesting to probe its effect on the visibility. In Fig.~\ref{fig:JCMomgk} we show the variation of visibility $\mathcal{V}$ with $\lambda$ for different values of $\omega$ and for fixed values of $\Delta \tau = 1$ and $\Delta E=1$. We see a periodic structure in this variation -- the periodicity of which depends on $\omega$. In the small $\lambda$ regime, i.e. for low noise, we see that the visibility is lower for higher values of $\omega$. 
\subsection{Bosonic field in a thermal state}

A more realistic case occurs when a thermal environment acts on the clock as it evolves inside a cavity. This translates to the environment being in a thermal state initially. In congruence with the earlier case of the environment being in a pure state, we assume that the cavity supports only one bosonic mode. In this case, a thermal state of the bath is a mixture of fock states corresponding to the single field mode, $\rho_{\beta}=\sum_n P_{\beta}(n)\ket{n}\bra{n}$, where $|n\rangle$ is the fock state basis of the field, and $\beta$ is the inverse temperature $\beta=1/K_B T$ such that $P_{\beta} (n) = \frac{e^{-n\omega/K_B T}}{\sum_n e^{-n\omega/K_B T}} \equiv \frac{e^{-n\beta\omega}}{1-e^{-\beta \omega}}$, which is representative of Boltzmann distribution.

To calculate the visibility, we consider the generalisation of Eq.~\eqref{eq:diff} to mixed states, which is $\mathcal{V}=|\trace \rho_{\textrm{clock}}\: e^{-iH_{\textrm{int}} \Delta\tau}|$ \cite{pikovski2015}. Therefore, the visibility in the case of a thermal state of the bosonic field is as follows:
\begin{align}
\mathcal{V}_{\beta}&=\bigg|\sum_n P_{\beta}(n) \bra{\psi_{n}} e^{-iH_{\textrm{int}} \Delta\tau}\ket{\psi_{n}}\bigg|\nonumber\\
&\equiv  \bigg|\sum_n P_{\beta}(n) \bra{\psi_{n}} U_\textrm{int}( \Delta\tau)\ket{\psi_{n}}\bigg|~,
\label{eq:thermo}
\end{align}
where $\ket{\psi_{n}} =  |\psi\rangle_\textrm{CE} (t=0) = |\psi_0\rangle \otimes |n\rangle_E$ is the initial extended state of the clock with environment in the $n$-th Fock state.

\begin{figure*}
  \centering
    \includegraphics[width=2\columnwidth]{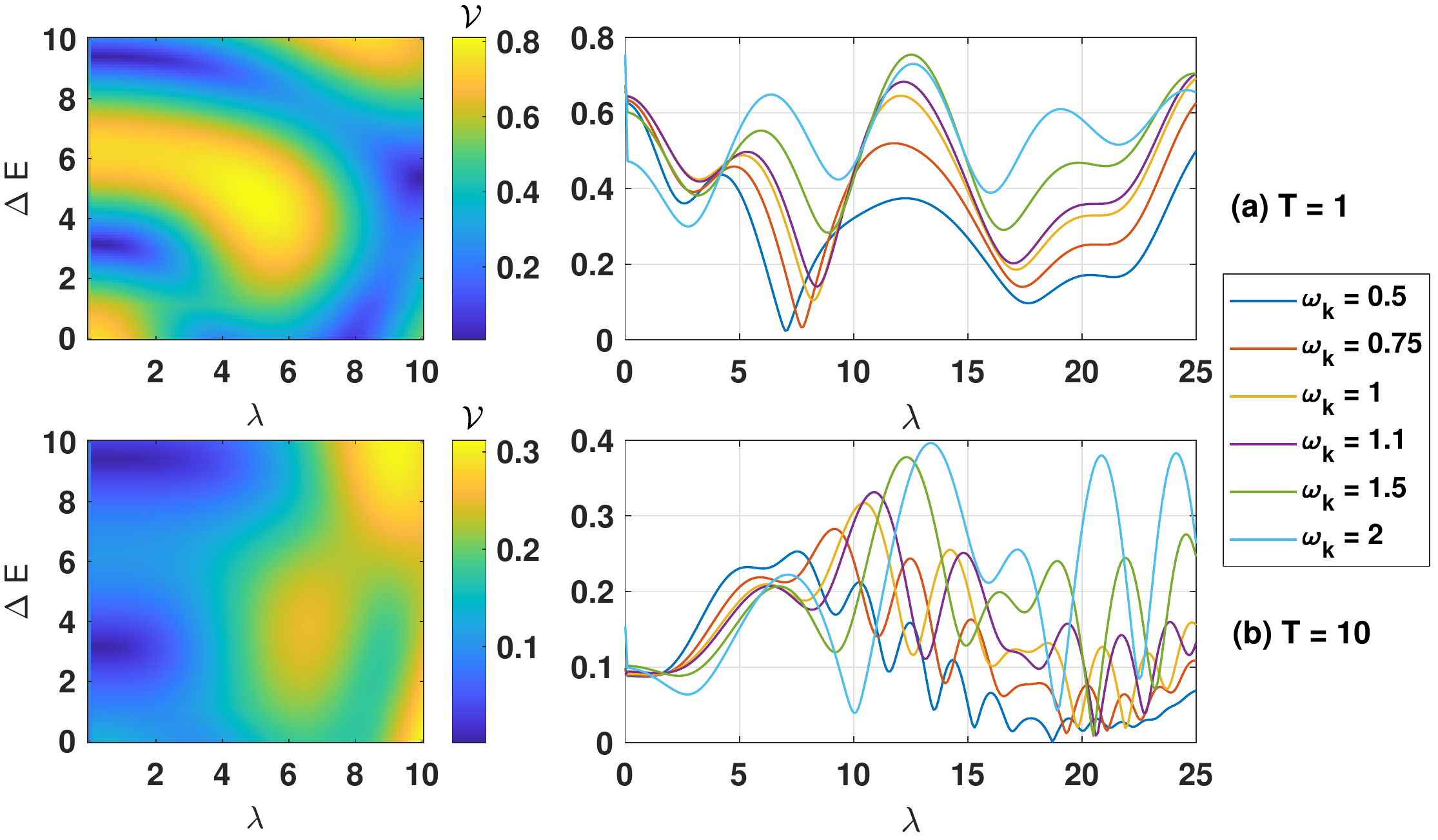}
    \caption{JC modelled environment interaction with finite temperature of the field. 2 cases are considered: (a) $T = 1$ -- low temperature, and (b) $T = 10$ -- high temperature. The left panel shows the variation of visibility on the clock energy difference ($\Delta E$) and the noise parameter $\lambda$ with fixed $\Delta \tau = 1$, $\omega = 1.1$. The right panel shows that the visibility also depends on the field frequency $\omega$ with fixed $\Delta E = 1$, $\Delta \tau = 1$. }
    \label{fig:JC_temp}
\end{figure*}

Using the general solution of the single mode Jaynes-Cummings model  \cite{JCmodel}, we provide the following eigenvectors:
\begin{align}
|v^1\rangle &= |00\rangle,\nonumber\\
|v_p^2\rangle &= \cos{\big(\theta_p\big)}|0,p+1\rangle - \sin{\big(\theta_p\big)}|1,p\rangle,\nonumber\\
|v_p^3\rangle &= \sin{\big(\theta_p\big)}|0,p+1\rangle + \cos{\big(\theta_p\big)}|1,p\rangle,
\end{align}
 where $|p\rangle$ denotes the excited cavity mode in the number basis. Also, $\cos(\theta_p) =  \frac{\Delta_p -\delta}{\sqrt{(\Delta_p - \delta)^2 + \Omega_p^2}}$, $\sin(\theta_p) =  \frac{\Omega_p}{\sqrt{(\Delta_p - \delta)^2 + \Omega_p^2}}$, $\Omega_p = \lambda\sqrt{p+1}$, and $\Delta_p = \sqrt{\delta^2+ \Omega_p^2}$,  and $\delta = \Delta E-\omega$ is the detuning frequency. The eigenvalues corresponding to the above eigenvectors are as follows:
\begin{align}
\mathcal{E}^{1}_p &= -\frac{\Delta E}{2}~,\nonumber\\
\mathcal{E}^{2}_p &= \bigg(p+\frac{1}{2}\bigg)\omega +\frac{\Delta_p}{2}~,\nonumber\\
\mathcal{E}^{3}_p &= \bigg(p+\frac{1}{2}\bigg)\omega -\frac{\Delta_p}{2}~.
\end{align}
For obtaining the visibility in Eq.~\eqref{eq:thermo}, the initial state of the combined system is expressed as: $|\psi\rangle_\textrm{CE} (t=0) = \bigg( \frac{|0\rangle + |1\rangle}{\sqrt{2}}\bigg) \otimes |n\rangle = \frac{1}{\sqrt{2}}\big[|0,n\rangle+|1,n\rangle\big]$. We express this initial state in terms of the above eigenvectors (renormalized following a basis change) as follows:
\begin{align}
&|\psi_n\rangle (0) = \frac{1}{\sqrt{2\mathcal{N}}}\bigg[ \big(\cos(\theta_{n-1}) |v^2_{n-1}\rangle-\sin(\theta_{n})|v^2_n\rangle\big) \nonumber\\
&+ \big(\sin(\theta_{n-1}) |v^3_{n-1}\rangle+\cos(\theta_{n})|v^3_n\rangle\big) \bigg]~,
\end{align}
with the normalization factor $\mathcal{N} =1$.
The time evolved state $|\psi_n\rangle (t) = U_{\textrm{int}}|\psi_n\rangle(0)$ is therefore written as follows:
\begin{align}
&|\psi_n\rangle (t) = \frac{1}{\sqrt{2}}\bigg[e^{-i\mathcal{E}^{2}_{n-1}t}\cos(\theta_{n-1})|v^2_{n-1}\rangle \nonumber\\
&- e^{-i\mathcal{E}^{2}_n t}\sin(\theta_n)|v^2_{n}\rangle+ e^{-i\mathcal{E}^{3}_{n-1}t}\sin(\theta_{n-1})|v^3_{n-1}\rangle \nonumber\\
&+ e^{-i\mathcal{E}^{3}_n t}\cos(\theta_n)|v^3_{n}\rangle\bigg]~.
\end{align}
With the exception of $n=0$, the above holds true for $n=1,2, ...\infty$, and we evaluate $\langle \psi_n(0)|U_\textrm{int}(\Delta\tau)|\psi_n(0) \rangle(\equiv \langle\psi_n (0)| \psi_n (\Delta\tau)\rangle)$ in Eq.~\eqref{eq:thermo} and find the following:
\begin{widetext}
\begin{align}
\langle\psi_n (0)| \psi_n (\Delta\tau)\rangle = \frac{1}{2} \bigg[e^{-i\mathcal{E}^{2}_{n-1}\Delta\tau}\cos^2(\theta_{n-1}) + e^{-i\mathcal{E}^{2}_n\Delta\tau}\sin^2(\theta_n)+ e^{-i\mathcal{E}^{3}_{n-1}\Delta\tau}\sin^2(\theta_{n-1}) + e^{-i\mathcal{E}^{3}_n\Delta\tau}\cos^2(\theta_n)\bigg]~.
\end{align}
\end{widetext}
This allows us to calculate the visibility in conjugation with the expression obtained for the case of $n=0$, which is mentioned in Eq.~\eqref{eq:JCstate}. 

While $n=0$ is the case with the effective temperature $T = 0$, we consider two more cases of $T = 1$ (low temperature), and $T = 10$ (high temperature). We illustrate the variation of visibility with $\Delta E$ and $\lambda$ (similar to Fig.~\eqref{fig:JCM}), and its dependence on $\omega$ (similar to Fig.~\eqref{fig:JCMomgk}) for these finite temperature cases in Fig.~\eqref{fig:JC_temp}. In general, it can be noted from the figure that  the visibility is in much lower range at high temperatures than at low temperatures as shown in the left panels on Fig.~\eqref{fig:JC_temp}. The right panel shows that the field frequency $\omega$ also controls the visibility. The coevolution of the environment and clock is evident and is generically driven by the complex interplay between noise and the intrinsic Hamiltonian of the clock and field.

\subsection{Visibility with ``clock in a cavity"}
Assuming that the JC model is effective in describing the dynamics of noisy clock, we quantify below the change in visibility on the basis of the calculations done in the last section.

To provide an estimate on the visibility in an actual experimental setting, we consider the earth's gravitational field which provides the proper time difference (to the first order in $c^2$) $\Delta \tau = \frac{g \Delta x t}{c^2}$. This is  $ \approx 10^{-15} s$ if we take the records for largest separable distance of a coherent superposition $\approx 0.5 m$ \cite{macrosup}, and coherence time $\approx 20 s$ \cite{Xu745}. If the energy scale is defined by the transition frequencies of atoms ($\omega\approx 10^{15}$ \cite{Zych2011}) which are commonly used in such experiments, then $\Delta \tau \Delta E \approx 1 $. If the above parameters can be achieved in the same experiment, then our calculations with $\Delta \tau = \Delta E =1$ are valid. Given that the experiment may be conducted in a cavity with some thermal environment despite cooling, our results of the JC model with the temperature $T=0.1$ hold significance as this would correspond to an effective temperature of $10^{-3} K$ i.e. in the $\rm mK$ range ($T=K_B T_\textrm{effective} \implies T_\textrm{effective} = 10^{-2} T$, assuming $K_B = 10^{2}$ in the scale defined by energy $\Delta E$), which is typical in dilation fridges available today. However, temperatures in the $nK$ range have been recently reported \cite{Xu745}. Clearly, this temperature is quite close to $T=0$ case which we considered. In the strong coupling regime i.e. for $\lambda =1$, as seen from Fig.~\eqref{fig:cc}, the difference between visibility without noise (0.8525) and that with the noise (0.7999) is $0.0526$, which is small.

\section{Clocks interacting with noise modelled on standard quantum channels}

In the theory of quantum computation, environmental effects giving rise to noise are modelled as quantum channels \cite{chuang,open}, that is to say, a Completely Positive and
Trace Preserving (CPTP) map acting on the system. It
is therefore interesting see how this type of noise would
affect the interference of clocks.

We will consider some noise models for qubits, which are based on the standard quantum channels -- AD, PD, and DP. Such models can be used to represent a wide variety of physical phenomena such as decay of a two-level system, dephasing etc. In this formalism, it is naturally assumed that the interaction of the environment with the system is for a finite time, say $\tau^*$. During this time physical interactions take place which can be associated with their corresponding effective probabilities, calculated after the interaction time $\tau^*$. Hence, the resultant density matrix at time $\tau^*$ is embedded with the information of these processes and their associated probabilities.

Continuous time models of the aforementioned quantum channels are usually described in the studies of open quantum systems with the time evolution of the system described by a Master equation. Due to the reasons explained in an earlier section, without resorting to the Master equation approach, to capture the effects of time dilation we wish to extend these finite-time models to continuous time evolution.

Here the system interacts with the environment through the Hamiltonian $H_{\textrm{noise}}$, and such that the original model is recovered if the time of interaction is set to $\tau^*$. The internal Hamiltonian governing the dynamics of the extended system, including the intrinsic internal, noiseless evolution of the clock is $H_{\textrm{int}} = H_0 + H_{\textrm{noise}}$. Recall that the internal Hamiltonian is defined in the reference frame of the system. Therefore, all references to time imply proper time along the system's semi-classical path, which generally differs from the time in the frame of reference of the laboratory due to time dilation. To get $H_{\textrm{noise}}$, we shall first find the time evolution operator of the extended system (clock + environment) corresponding to the noise which is based on quantum channels. We can then work backwards to obtain the Hamiltonian $H_\textrm{noise}$ knowing that the time evolution operator corresponding to the noise acts for finite time $\tau^*$. Then, by including the intrinsic evolution of the two-level system, i.e.\ the clock, we evolve the initial state of the extended system given in Eq.~\eqref{eq:init} through a unitary operator which captures the natural dynamics of the system as well as the noise ($U_\textrm{int}$) and then we calculate the visibility by taking the scalar product of the final states (Eq.~\eqref{eq:diff}). 

In the appendix, we consider noise models based on standard quantum channels one by one, provide the time evolution operators of the extended system and calculate the visibility subject to various parameters such as the difference in proper times through the arms of the interferometer ($\Delta\tau$), the energy scale of the interactions in various noise models and the difference in energy levels of the two level system which constitutes a clock ($\Delta E$). For simplicity, here we consider the minimal extension to the system, i.e., the smallest dimensionality of the environment which can capture the qualitative aspects of the adhered noise model.

For capturing the qualitative effects of noise, particularly in the practical limit of small proper time difference ($\Delta \tau$), a finite size environment is adequate to be modelled as interacting with the clock. However, owing to the finite size of the environment, we can expect to see recurrences in the dynamics \cite{Bocchieri1957} of the system, which might not be necessarily relevant in every experiment given that they would only assume importance in much longer time-scales than $\Delta\tau$. Therefore, despite this peculiarity, as long as the time of evolution (proper time) of the clock is small relative to the recurrence times, this model would remain appropriate in an experimental setting and thus, qualitatively sound for analyzing the environmental effects as intended. That being said, the effects of a small-dimensional environment are conceptually interesting in their own right \cite{BAN2021126936} and might be relevant for studying non-Markovian dynamics.

\section{Conclusions}

Clock interferometry is one of the promising theoretical proposals which captures the interplay between quantum mechanics and gravity. The drop in interferometric visibility can be seen as a genuine effect of time dilation due to gravity on a quantum system. The clocks considered in a wide variety of physical implementations of the proposal would have susceptibility to noise which would, in turn, affect the visibility. Therefore, the focus of this manuscript has been on noise affecting the clock, which assumes importance as it only manifests in effect due to different proper times along the paths in a MZ setup. We have established that the interplay between the extension of Hilbert space due to modelling of the environment as additional DOFs interacting with the clock and corruption of which path information by the noise leads to a change in the interferometric visibility.

At first, we probed the effect of noise on the internal DOF in a very generic setting without going into the specifics of any experimental realization or noise model and established that the form of visibility remains the same, albeit one has to consider the state of both system and environment to be time evolved in a time-dilated setting within the arms of interferometer. Intrinsically, our model incorporates the effect of time dilation on the environmental interaction as well.

By considering a simple model of a single mode bosonic field acting as the environment we ascertained the effect of noise on the internal DOF and the visibility. To analyse the effect of finite temperature conditions encountered by a clock located inside a cavity, we considered the JC model with finite temperature environment and found that a higher temperature leads to a lower visibility for a small amount of noise, and small proper time difference. Additionally, we probed the noisy evolution of a clock by using noise models based on standard quantum channels. In all of these cases, we established that the environmental effect is contingent on the energy scale of the Hamiltonian representing the interaction of the clocks with the environment, apart from the clock Hamiltonian.


In the quantum channel based noise models, we specifically investigated the fate of the visibility subject to change in energy levels of the clock, proper time difference and the energy scale of noise Hamiltonian. The visibilities for all types of noise, show different susceptibilities to the energy scale of noise Hamiltonian. Our results for the channel based noise models  are in strong agreement with the intuitive inferences drawn in the low noise regime. In general, one may conclude that the visibility may decrease or even increase in the presence of an environment due to a complex behaviour arising out of the aforementioned competing effects and their interplay. As a peculiar case, in channel based noise models we also considered the scenario where the noise acts differently on the two arms, providing additional which-way information besides that offered by the clock --- and thus a distinct source of decoherence. Therefore, in this case, even for no proper time difference, we saw a decrease in the visibility.

We also quantified the change in visibility purely due to environmental noise (based on JC model) while taking into consideration best available parameters from distinct experiments. It is notable that this quantification is only valid in particular experimental conditions -- summarily mentioned as ``clock in a cavity". This effect of noise on visibility, though small, establishes that one cannot disregard noise on internal DOF in probing the effect of gravity. As the coherence times, and/or the separation between the arms of the interferometer increase, the effect of gravity will be more easy to detect and we expect to see a substantial difference in the observed visibility stemming from a noisy clock. Recent advances in quantum technology are reassuring towards achievement of the ``best" parameters which are taken from different experiments in a single experiment in the near future.

\acknowledgments
M.Z and F.C.~acknowledge support through Australian Research Council (ARC) DECRA grants DE180101443 and DE170100712, and ARC Centre EQuS CE170100009. The authors acknowledge the traditional owners of the land on which the University of Queensland is situated, the Turrbal and Jagera people.

\printbibliography

@article{Bocchieri1957,
  title = {Quantum Recurrence Theorem},
  author = {Bocchieri, P. and Loinger, A.},
  journal = {Phys. Rev.},
  volume = {107},
  issue = {2},
  pages = {337--338},
  numpages = {0},
  year = {1957},
  month = {Jul},
  publisher = {American Physical Society},
  doi = {10.1103/PhysRev.107.337},
  url = {https://link.aps.org/doi/10.1103/PhysRev.107.337}
}

@Article{Zych2011,
author={Zych, Magdalena
and Costa, Fabio
and Pikovski, Igor
and Brukner, Caslav},
title={Quantum interferometric visibility as a witness of general relativistic proper time},
journal={Nature Communications},
year={2011},
month={Oct},
day={18},
publisher={Nature Publishing Group},
volume={2},
pages={505},
doi = {10.1038/ncomms1498},
url={10.1038/ncomms1498}
}

@article{Zych2016,
	doi = {10.1088/1742-6596/723/1/012044},
	url = {10.1088/1742-6596/723/1/012044},
	year = 2016,
	month = {jun},
	publisher = {{IOP} Publishing},
	volume = {723},
	pages = {012044},
	author = {M Zych and I Pikovski and F Costa and {\v{C}} Brukner},
	title = {General relativistic effects in quantum interference of {\textquotedblleft}clocks{\textquotedblright}},
	journal = {Journal of Physics: Conference Series}
}

@article{neutro1,
  title = {Observation of Gravitationally Induced Quantum Interference},
  author = {Colella, R. and Overhauser, A. W. and Werner, S. A.},
  journal = {Phys. Rev. Lett.},
  volume = {34},
  issue = {23},
  pages = {1472--1474},
  numpages = {0},
  year = {1975},
  month = {Jun},
  publisher = {American Physical Society},
  doi = {10.1103/PhysRevLett.34.1472},
  url = {https://link.aps.org/doi/10.1103/PhysRevLett.34.1472}
}

@article{complement,
  title = {Fringe Visibility and Which-Way Information: An Inequality},
  author = {Englert, Berthold-Georg},
  journal = {Phys. Rev. Lett.},
  volume = {77},
  issue = {11},
  pages = {2154--2157},
  numpages = {0},
  year = {1996},
  month = {Sep},
  publisher = {American Physical Society},
  doi = {10.1103/PhysRevLett.77.2154},
  url = {https://link.aps.org/doi/10.1103/PhysRevLett.77.2154}
}

@article{pikovski2015,
title={Universal decoherence due to gravitational time dilation},
author={Pikovski, Igor and Zych, Magdalena and Costa, Fabio and Brukner, {\v{C}}aslav},
journal={Nature Physics},
year={2015},
month={Jun},
publisher={Nature Publishing Group},
volume={11},
pages={668},
url= {https://doi.org/10.1038/nphys3366},
doi= {10.1038/nphys3366}
}

@article{pikovski2017,
  title={Time dilation in quantum systems and decoherence},
  author={Pikovski, Igor and Zych, Magdalena and Costa, Fabio and Brukner, {\v{C}}aslav},
  journal={New Journal of Physics},
  volume={19},
  number={2},
  pages={025011},
  year={2017},
  publisher={IOP Publishing},
  doi = {10.1088/1367-2630/aa5d92},
  url = {10.1088}
}

@article{sat1,
	title = {Quantum optics experiments using the International Space Station: a proposal},
	author = {T Scheidl and E Wille and R Ursin},
	journal = {New Journal of Physics},
	volume = {15},
	number = {4},
	pages = {043008},
	doi = {10.1088/1367-2630/15/4/043008},
	url = {10.1088},
	year = 2013,
	month = {apr},
	publisher = {{IOP} Publishing},
}

@article{sat2,
  title = {Spacetime effects on satellite-based quantum communications},
  author = {Bruschi, David Edward and Ralph, Timothy C. and Fuentes, Ivette and Jennewein, Thomas and Razavi, Mohsen},
  journal = {Phys. Rev. D},
  volume = {90},
  issue = {4},
  pages = {045041},
  numpages = {13},
  year = {2014},
  month = {Aug},
  publisher = {American Physical Society},
  doi = {10.1103/PhysRevD.90.045041},
  url = {https://link.aps.org/doi/10.1103/PhysRevD.90.045041}
}

@article {Marg1205,
	author = {Margalit, Yair and Zhou, Zhifan and Machluf, Shimon and Rohrlich, Daniel and Japha, Yonathan and Folman, Ron},
	title = {A self-interfering clock as a {\textquotedblleft}which path{\textquotedblright} witness},
	volume = {349},
	number = {6253},
	pages = {1205--1208},
	year = {2015},
	doi = {10.1126/science.aac6498},
	publisher = {American Association for the Advancement of Science},
	issn = {0036-8075},
	URL = {https://science.sciencemag.org/content/349/6253/1205},
	journal = {Science}
}

@article{macrosup, 
author={Kovachy, T. and Asenbaum, P. and Overstreet, C. and Donnelly, C. A. and Dickerson, S. M.
and Sugarbaker, A.
and Hogan, J. M.
and Kasevich, M. A.},
title={Quantum superposition at the half-metre scale},
journal={Nature},
year={2015},
month={Dec},
day={23},
publisher={Nature Publishing Group},
volume={528},
pages={530},
url={https://doi.org/10.1038/nature16155},
doi= {10.1038/nature16155}
}

@article{singleE,
	doi = {10.1088/1367-2630/aa638f},
	url = {https://doi.org/10.1088},
	year = 2017,
	month = {mar},
	publisher = {{IOP} Publishing},
	volume = {19},
	number = {3},
	pages = {033028},
	author = {Christopher Hilweg and Francesco Massa and Denis Martynov and Nergis Mavalvala and Piotr T Chru{\'{s}}ciel and Philip Walther},
	title = {Gravitationally induced phase shift on a single photon},
	journal = {New Journal of Physics}
	}

@article{singleP,
	doi = {10.1088/1367-2630/18/9/093050},
	url = {https://doi.org/10.1088},
	year = 2016,
	month = {sep},
	publisher = {{IOP} Publishing},
	volume = {18},
	number = {9},
	pages = {093050},
	author = {P A Bushev and J H Cole and D Sholokhov and N Kukharchyk and M Zych},
	title = {Single electron relativistic clock interferometer},
	journal = {New Journal of Physics}
	}

@article{Zych_2012,
	doi = {10.1088/0264-9381/29/22/224010},
	url = {https://doi.org/10.1088},
	year = 2012,
	month = {oct},
	publisher = {{IOP} Publishing},
	volume = {29},
	number = {22},
	pages = {224010},
	author = {Magdalena Zych and Fabio Costa and Igor Pikovski and Timothy C Ralph and {\v{C}}aslav Brukner},
	title = {General relativistic effects in quantum interference of photons},
	journal = {Classical and Quantum Gravity}
	}

@article {2019arxiv,
	author = {Loriani, Sina and Friedrich, Alexander and Ufrecht, Christian and Di Pumpo, Fabio and Kleinert, Stephan and Abend, Sven and Gaaloul, Naceur and Meiners, Christian and Schubert, Christian and Tell, Dorothee and Wodey, {\'E}tienne and Zych, Magdalena and Ertmer, Wolfgang and Roura, Albert and Schlippert, Dennis and Schleich, Wolfgang P. and Rasel, Ernst M. and Giese, Enno},
	title = {Interference of clocks: A quantum twin paradox},
	volume = {5},
	number = {10},
	elocation-id = {eaax8966},
	year = {2019},
	publisher = {American Association for the Advancement of Science},
	URL = {https://advances.sciencemag.org/content/5/10/eaax8966},
	journal = {Science Advances}
}

@article{sat3,
  title = {Interference at the Single Photon Level Along Satellite-Ground Channels},
  author = {Vallone, Giuseppe and Dequal, Daniele and Tomasin, Marco and Vedovato, Francesco and Schiavon, Matteo and Luceri, Vincenza and Bianco, Giuseppe and Villoresi, Paolo},
  journal = {Phys. Rev. Lett.},
  volume = {116},
  issue = {25},
  pages = {253601},
  numpages = {6},
  year = {2016},
  month = {Jun},
  publisher = {American Physical Society},
  doi = {10.1103/PhysRevLett.116.253601},
  url = {https://link.aps.org/doi/10.1103/PhysRevLett.116.253601}
}

@article{atom1,
  title = {Atom Interferometry with the Sr Optical Clock Transition},
  author = {Hu, Liang and Poli, Nicola and Salvi, Leonardo and Tino, Guglielmo M.},
  journal = {Phys. Rev. Lett.},
  volume = {119},
  issue = {26},
  pages = {263601},
  numpages = {5},
  year = {2017},
  month = {Dec},
  publisher = {American Physical Society},
  doi = {10.1103/PhysRevLett.119.263601},
  url = {https://link.aps.org/doi/10.1103/PhysRevLett.119.263601}
}

@article{neutro2,
  title = {Gravity and inertia in quantum mechanics},
  author = {Staudenmann, J. -L. and Werner, S. A. and Colella, R. and Overhauser, A. W.},
  journal = {Phys. Rev. A},
  volume = {21},
  issue = {5},
  pages = {1419--1438},
  numpages = {0},
  year = {1980},
  month = {May},
  publisher = {American Physical Society},
  doi = {10.1103/PhysRevA.21.1419},
  url = {https://link.aps.org/doi/10.1103/PhysRevA.21.1419}
}

@article{unstable,
title = "A priori which-way information in quantum interference with unstable particles",
journal = "Physics Letters A",
volume = "378",
number = "34",
pages = "2490 - 2494",
year = "2014",
issn = "0375-9601",
doi = "10.1016/j.physleta.2014.06.036",
url = "http://www.sciencedirect.com/science/article/pii/S037596011400632X",
author = "D.E. Krause and E. Fischbach and Z.J. Rohrbach"
}

@article{atom2,
  title = {Relativistic effects in atom and neutron interferometry and the differences between them},
  author = {Greenberger, Daniel M. and Schleich, Wolfgang P. and Rasel, Ernst M.},
  journal = {Phys. Rev. A},
  volume = {86},
  issue = {6},
  pages = {063622},
  numpages = {16},
  year = {2012},
  month = {Dec},
  publisher = {American Physical Society},
  doi = {10.1103/PhysRevA.86.063622},
  url = {https://link.aps.org/doi/10.1103/PhysRevA.86.063622}
}

@article{Zhou_2018,
	doi = {10.1088/1361-6382/aad56b},
	url = {https://doi.org/10.1088/1361-6382/aad56b},
	year = 2018,
	month = {aug},
	publisher = {{IOP} Publishing},
	volume = {35},
	number = {18},
	pages = {185003},
	author = {Zhifan Zhou and Yair Margalit and Daniel Rohrlich and Yonathan Japha and Ron Folman},
	title = {Quantum complementarity of clocks in the context of general relativity},
	journal = {Classical and Quantum Gravity}
}

@article{Carney_2019,
	doi = {10.1088/1361-6382/aaf9ca},
	url = {https://doi.org/10.1088/1361-6382/aaf9ca},
	year = 2019,
	month = {jan},
	publisher = {{IOP} Publishing},
	volume = {36},
	number = {3},
	pages = {034001},
	author = {Daniel Carney and Philip C E Stamp and Jacob M Taylor},
	title = {Tabletop experiments for quantum gravity: a user's manual},
	journal = {Classical and Quantum Gravity}
	}

@article{Bose,
  title = {Spin Entanglement Witness for Quantum Gravity},
  author = {Bose, Sougato and Mazumdar, Anupam and Morley, Gavin W. and Ulbricht, Hendrik and Toro\ifmmode \check{s}\else \v{s}\fi{}, Marko and Paternostro, Mauro and Geraci, Andrew A. and Barker, Peter F. and Kim, M. S. and Milburn, Gerard},
  journal = {Phys. Rev. Lett.},
  volume = {119},
  issue = {24},
  pages = {240401},
  numpages = {6},
  year = {2017},
  month = {Dec},
  publisher = {American Physical Society},
  doi = {10.1103/PhysRevLett.119.240401},
  url = {https://link.aps.org/doi/10.1103/PhysRevLett.119.240401}
}

@article{Vedral,
  title = {Gravitationally Induced Entanglement between Two Massive Particles is Sufficient Evidence of Quantum Effects in Gravity},
  author = {Marletto, C. and Vedral, V.},
  journal = {Phys. Rev. Lett.},
  volume = {119},
  issue = {24},
  pages = {240402},
  numpages = {5},
  year = {2017},
  month = {Dec},
  publisher = {American Physical Society},
  doi = {10.1103/PhysRevLett.119.240402},
  url = {https://link.aps.org/doi/10.1103/PhysRevLett.119.240402}
}

@article{Markus,
  title = {Quantum superposition of massive objects and the quantization of gravity},
  author = {Belenchia, Alessio and Wald, Robert M. and Giacomini, Flaminia and Castro-Ruiz, Esteban and Brukner, \ifmmode \check{C}\else \v{C}\fi{}aslav and Aspelmeyer, Markus},
  journal = {Phys. Rev. D},
  volume = {98},
  issue = {12},
  pages = {126009},
  numpages = {9},
  year = {2018},
  month = {Dec},
  publisher = {American Physical Society},
  doi = {10.1103/PhysRevD.98.126009},
  url = {https://link.aps.org/doi/10.1103/PhysRevD.98.126009}
}

@article{AB,
  title = {Significance of Electromagnetic Potentials in the Quantum Theory},
  author = {Aharonov, Y. and Bohm, D.},
  journal = {Phys. Rev.},
  volume = {115},
  issue = {3},
  pages = {485--491},
  numpages = {0},
  year = {1959},
  month = {Aug},
  publisher = {American Physical Society},
  doi = {10.1103/PhysRev.115.485},
  url = {https://link.aps.org/doi/10.1103/PhysRev.115.485}
}

@article{JCmodel, 
author={E. T. {Jaynes} and F. W. {Cummings}}, 
journal={Proceedings of the IEEE}, 
title={Comparison of quantum and semiclassical radiation theories with application to the beam maser}, 
year={1963}, 
volume={51}, 
number={1}, 
pages={89-109}, 
doi={10.1109/PROC.1963.1664}, 
ISSN={0018-9219}, 
month={Jan},}

@book{chuang, place={Cambridge}, title={Quantum Computation and Quantum Information: 10th Anniversary Edition}, DOI={10.1017/CBO9780511976667}, publisher={Cambridge University Press}, author={Nielsen, Michael A. and Chuang, Isaac L.}, year={2010}}

@book{open, place={Cambridge}, title={The Mathematical Language of Quantum Theory: From Uncertainty to Entanglement}, DOI={10.1017/CBO9781139031103}, publisher={Cambridge University Press}, author={Heinosaari, Teiko and Ziman, Mário}, year={2011}}

@Article{Fein2019,
author={Fein, Yaakov Y.
and Geyer, Philipp
and Zwick, Patrick
and Kialka, Filip
and Pedalino, Sebastian
and Mayor, Marcel
and Gerlich, Stefan
and Arndt, Markus},
title={Quantum superposition of molecules beyond 25 kDa},
journal={Nature Physics},
year={2019},
issn={1745-2481},
url={https://doi.org/10.1038/s41567-019-0663-9},
doi={10.1038/s41567-019-0663-9}
}

@article {Xu745,
	author = {Xu, Victoria and Jaffe, Matt and Panda, Cristian D. and Kristensen, Sofus L. and Clark, Logan W. and M{\"u}ller, Holger},
	title = {Probing gravity by holding atoms for 20 seconds},
	volume = {366},
	number = {6466},
	pages = {745--749},
	year = {2019},
	publisher = {American Association for the Advancement of Science},
	issn = {0036-8075},
	URL = {https://science.sciencemag.org/content/366/6466/745},
	journal = {Science}
}

@article{BAN2021126936,
title = "Two-qubit correlation in two independent environments with indefiniteness",
journal = "Physics Letters A",
volume = "385",
pages = "126936",
year = "2021",
issn = "0375-9601",
doi = "10.1016/j.physleta.2020.126936",
url = "http://www.sciencedirect.com/science/article/pii/S0375960120308033",
author = "Masashi Ban",
}

@Article{Krisnanda2020,
author={Krisnanda, Tanjung
and Tham, Guo Yao
and Paternostro, Mauro
and Paterek, Tomasz},
title={Observable quantum entanglement due to gravity},
journal={npj Quantum Information},
year={2020},
month={Jan},
day={30},
volume={6},
number={1},
pages={12},
doi={10.1038/s41534-020-0243-y},
url={https://doi.org/10.1038/s41534-020-0243-y}
}

@article{Bose2,
  title = {Locality and entanglement in table-top testing of the quantum nature of linearized gravity},
  author = {Marshman, Ryan J. and Mazumdar, Anupam and Bose, Sougato},
  journal = {Phys. Rev. A},
  volume = {101},
  issue = {5},
  pages = {052110},
  numpages = {13},
  year = {2020},
  month = {May},
  publisher = {American Physical Society},
  doi = {10.1103/PhysRevA.101.052110},
  url = {https://link.aps.org/doi/10.1103/PhysRevA.101.052110}
}

@article{unknown,
  title = {Witnessing the nonclassical nature of gravity in the presence of unknown interactions},
  author = {Chevalier, Hadrien and Paige, A. J. and Kim, M. S.},
  journal = {Phys. Rev. A},
  volume = {102},
  issue = {2},
  pages = {022428},
  numpages = {10},
  year = {2020},
  month = {Aug},
  publisher = {American Physical Society},
  doi = {10.1103/PhysRevA.102.022428},
  url = {https://link.aps.org/doi/10.1103/PhysRevA.102.022428}
}

@article{Guerreiro_2020,
	doi = {10.1088/1361-6382/ab9d5d},
	url = {https://doi.org/10.1088/1361-6382/ab9d5d},
	year = 2020,
	month = {jul},
	publisher = {{IOP} Publishing},
	volume = {37},
	number = {15},
	pages = {155001},
	author = {Thiago Guerreiro},
	title = {Quantum effects in gravity waves},
	journal = {Classical and Quantum Gravity}
}

@Article{Asenbaum2017,
  author    = {Asenbaum, Peter and Overstreet, Chris and Kovachy, Tim and Brown, Daniel D. and Hogan, Jason M. and Kasevich, Mark A.},
  journal   = {Phys. Rev. Lett.},
  title     = {Phase Shift in an Atom Interferometer due to Spacetime Curvature across its Wave Function},
  year      = {2017},
  month     = {May},
  pages     = {183602},
  volume    = {118},
  doi       = {10.1103/PhysRevLett.118.183602},
  issue     = {18},
  numpages  = {5},
  publisher = {American Physical Society},
  url       = {https://link.aps.org/doi/10.1103/PhysRevLett.118.183602},
}

@article{dephase,
  title = {Entanglement, Dephasing, and Phase Recovery via Cross-Correlation Measurements of Electrons},
  author = {Neder, I. and Heiblum, M. and Mahalu, D. and Umansky, V.},
  journal = {Phys. Rev. Lett.},
  volume = {98},
  issue = {3},
  pages = {036803},
  numpages = {4},
  year = {2007},
  month = {Jan},
  publisher = {American Physical Society},
  doi = {10.1103/PhysRevLett.98.036803},
  url = {https://link.aps.org/doi/10.1103/PhysRevLett.98.036803}
}

@article{Westphal2021,
author={Westphal, Tobias
and Hepach, Hans
and Pfaff, Jeremias
and Aspelmeyer, Markus},
title={Measurement of gravitational coupling between millimetre-sized masses},
journal={Nature},
year={2021},
month={Mar},
day={01},
volume={591},
number={7849},
pages={225-228},
issn={1476-4687},
doi={10.1038/s41586-021-03250-7},
url={https://doi.org/10.1038/s41586-021-03250-7}
}
\onecolumn\newpage
\appendix
\section{Derivation and analysis of Visibility for standard quantum channels in the arms of the MZ interferometer}

\subsection{Amplitude Damping Channel}

An AD channel represents the process of spontaneous emission of a photon from the decay of excited state of an atom or molecule. In a finite temperature environment, a photon may also be captured by the atom in the ground state leading it to attain the excited state. We  consider here a two level system as a clock and use an additional qubit to represent the environment. However, the JC model with a single field mode, sec.~IV, better represents the photon emission process than the model with a single qubit as environment used here. To construct the unitary operator governing the evolution of the extended system, we consider a probabilistic process of absorption of photon by the clock along with the emission which is in contrast with the JC model, for which the photon recapture process is mitigated by the dynamics of the bosonic field.

In this single qubit environment, if the clock decays from its excited state ($|1\rangle_\textrm{C} \rightarrow |0\rangle_\textrm{C}$), the emission of the photon is represented by an excited environmental qubit ($|0\rangle_\textrm{E} \rightarrow |1\rangle_\textrm{E}$). In case that the clock is excited by the environment ($|0\rangle_\textrm{C} \rightarrow |1\rangle_\textrm{C}$), it subsequently leads to decay of the environmental qubit to its ground state ($|1\rangle_\textrm{E} \rightarrow |0\rangle_\textrm{E}$). Of course, each of these processes are probabilistic in nature. It turns out that the minimum dimensionality of the environment required for representing this behaviour is 2 as we have modelled above.

Keeping in mind the above probabilistic transitions, we construct the following unitary time evolution operator acting on the extended system: 
\begin{align}
U_\textrm{AD} (\tau^*)=
  \begin{bmatrix}
    1 & 0 & 0 & 0 \\
    0 & \sqrt{1-p} & \sqrt{p} & 0\\
    0 & -\sqrt{p} & \sqrt{1-p} & 0 \\
    0 & 0 & 0 & 1
  \end{bmatrix},
\end{align}
where $p$ represents the probability of transition to take place which is dependent on the time scale $\tau^*$ of interaction of the clock with environment. $U_\textrm{AD}$ has the eigenvalues : $\{1,1,\sqrt{1-p}-i\sqrt{p},\sqrt{1-p}+i\sqrt{p}\}$ with the eigenvectors as follows:
\begin{alignat}{2}
|v_1'\rangle=|00\rangle,~|v_2'\rangle=|11\rangle,~|v_3'\rangle= i|01\rangle +|10\rangle,~|v_4'\rangle= -i|01\rangle +|10\rangle .
\end{alignat}

On substituting $p=\sin^2\theta$, the eigenvalues become $\{1,1,e^{-i\theta},e^{i\theta}\}$. The eigenvalues of the corresponding Hamiltonian $H_\textrm{AD}$ governing the dynamics of the clock and environment through AD channel are: $\{0,0,-\theta/\tau^*,\theta/\tau^*\}$ where $\tau^*$ is the time scale of interaction as mentioned before. Its eigenvectors are the same as that of $U_\textrm{AD}$. Therefore, we write the Hamiltonian operator as follows:
\begin{align}\label{ADH}
H_\textrm{AD} = -\frac{\theta}{\tau^*}~|v_3'\rangle\langle v_3'| + \frac{\theta}{\tau^*}~|v_4'\rangle\langle v_4'|\equiv 2i\lambda\big[|10\rangle\langle 01|-|01\rangle\langle10|\big],
\end{align}
where $|v_3'\rangle$ and $|v_4'\rangle$ are the eigenvectors corresponding to non zero eigenvalues of $H_\textrm{AD}$ and $\lambda = \frac{\theta}{\tau^*} $ represents the energy scale of the interaction between the clock and environment. Apart from the aforementioned transitions, the extended system also evolves under its intrinsic  dynamics through a Hamiltonian $H_0$ which represents a two level system with the added ancillary qubit of the environment and which is given as follows:
\begin{align}
H_0 = \left(E_0 |0\rangle\langle 0| +E_1 |1\rangle\langle 1|\right)\otimes\mathbb{I}, 
\label{eq:H0}
\end{align}
where $E_0$ and $E_1$ are the energy levels of the two level system serving as a clock. Combining $H_{0}$ and $H_\textrm{AD}$, we get the total Hamiltonian of the extended system as follows: 
\begin{align}
H_\textrm{int}=H_{0}+H_\textrm{AD}
= \begin{bmatrix}
 E_0 & 0& 0& 0 \\
 0& E_0& -2 i \lambda& 0 \\
 0& 2 i \lambda& E_1& 0 \\
 0& 0& 0& E_1
\end{bmatrix}.
\end{align}
For the continuous time evolution of the extended system, we diagonalize the above Hamiltonian to obtain its eigenvalues and eigenvectors and hence, construct the corresponding time evolution operator. The initial state of the extended system,  $|\psi\rangle_\textrm{CE}(t=0)$ (Eq.~\eqref{eq:init}) can be expressed  in terms of the eigenvectors {$|v_1\rangle,|v_2\rangle,|v_3\rangle,|v_4\rangle$} of $H_\textrm{int}$ as:
 \begin{align}
 |\psi\rangle_\textrm{CE} (t=0)= \frac{1}{\sqrt2}\bigg[|v_1\rangle+\bigg(\frac{1-y}{2}\bigg)|v_3\rangle +\bigg(\frac{1+y}{2}\bigg)|v_4\rangle\bigg],
 \end{align}
where $y=\frac{\Delta E}{ \sqrt{\Delta E^2 +16\lambda^2}},\Delta E =E_1-E_0$. The eigenvalues of $H_\textrm{int}$ are \{$E_\textrm{0}, E_\textrm{1}, \frac{(E_+ - \Delta E/y)}{2}, \frac{(E_+ + \Delta E/y)}{2}$\}, where $E_+=E_0+E_1$. We can now construct the unitary time evolution operator for $H_\textrm{int}$ and apply it to the initial state (Eq.~\eqref{eq:init}). At a later time instant t, the clock + environment state $|\psi\rangle_\textrm{CE}(t) = U_\textrm{int} |\psi\rangle_\textrm{CE} (t=0) = e^{-i H_{\textrm{int}}t}  |\psi\rangle_\textrm{CE} (t=0)$is given as follows:
\begin{align}
|\psi\rangle_\textrm{CE}(t) =
           \frac{e^{-i E_0 t}}{\sqrt{2}}|v_1\rangle + \frac{(1-y)e^{-it\frac{(E_+ -\Delta E/y)}{2}}}{2\sqrt{2}} |v_3\rangle + \frac{(1+y)e^{-it\frac{(E_+ +\Delta E/y)}{2}}}{2\sqrt{2}} |v_4\rangle .
 \label{eq:ADstate}
 \end{align}

\begin{figure}
    \centering
    \includegraphics[width=0.5\columnwidth]{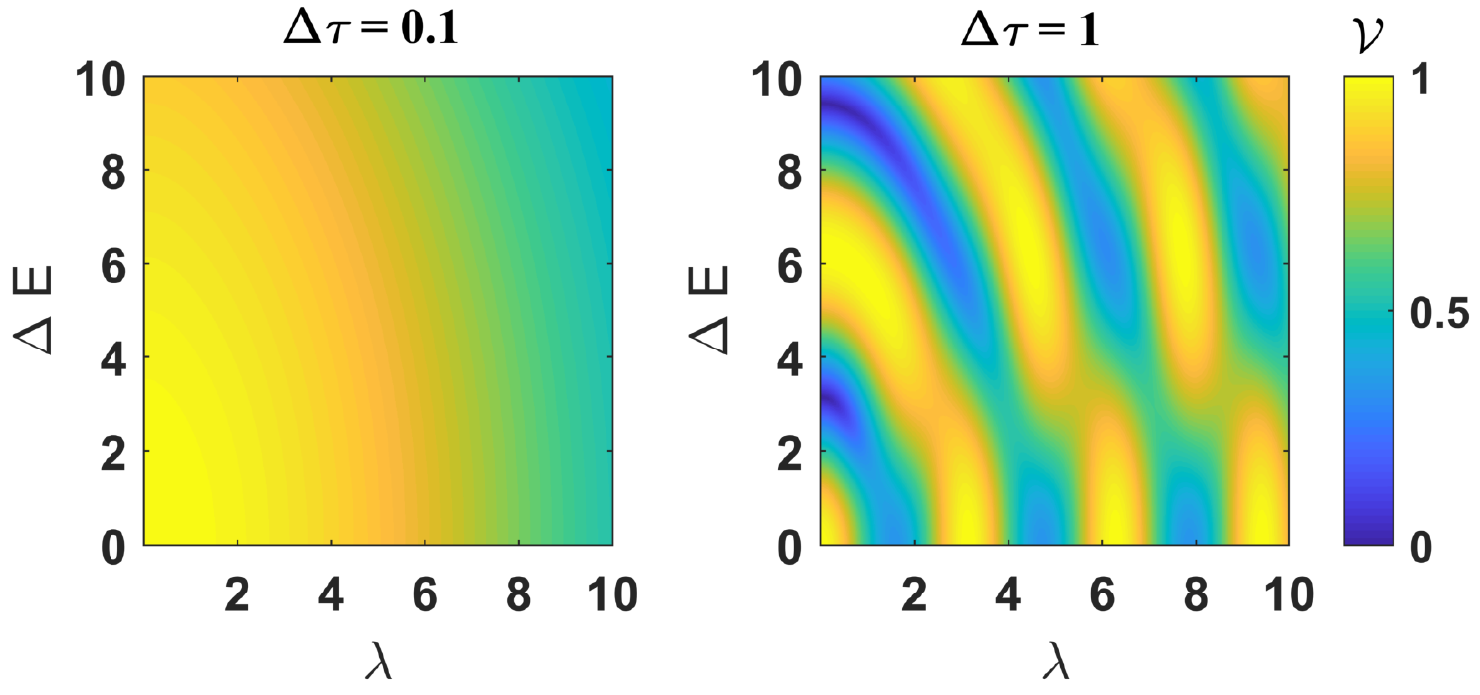}
   \caption{The variation of visibility ($\mathcal{V}$) with $\Delta E$ and noise parameter $\lambda$ in case of noise model based on AD channel, for different $\Delta \tau$. The interplay of these leads to formation of fringes in the visibility pattern. It is assumed that $\lambda_1=\lambda_2$. An independent change in either of these leads to oscillatory behaviour in the visibility.}
    \label{fig:AD4}
    \centering
    \includegraphics[width=0.5\columnwidth]{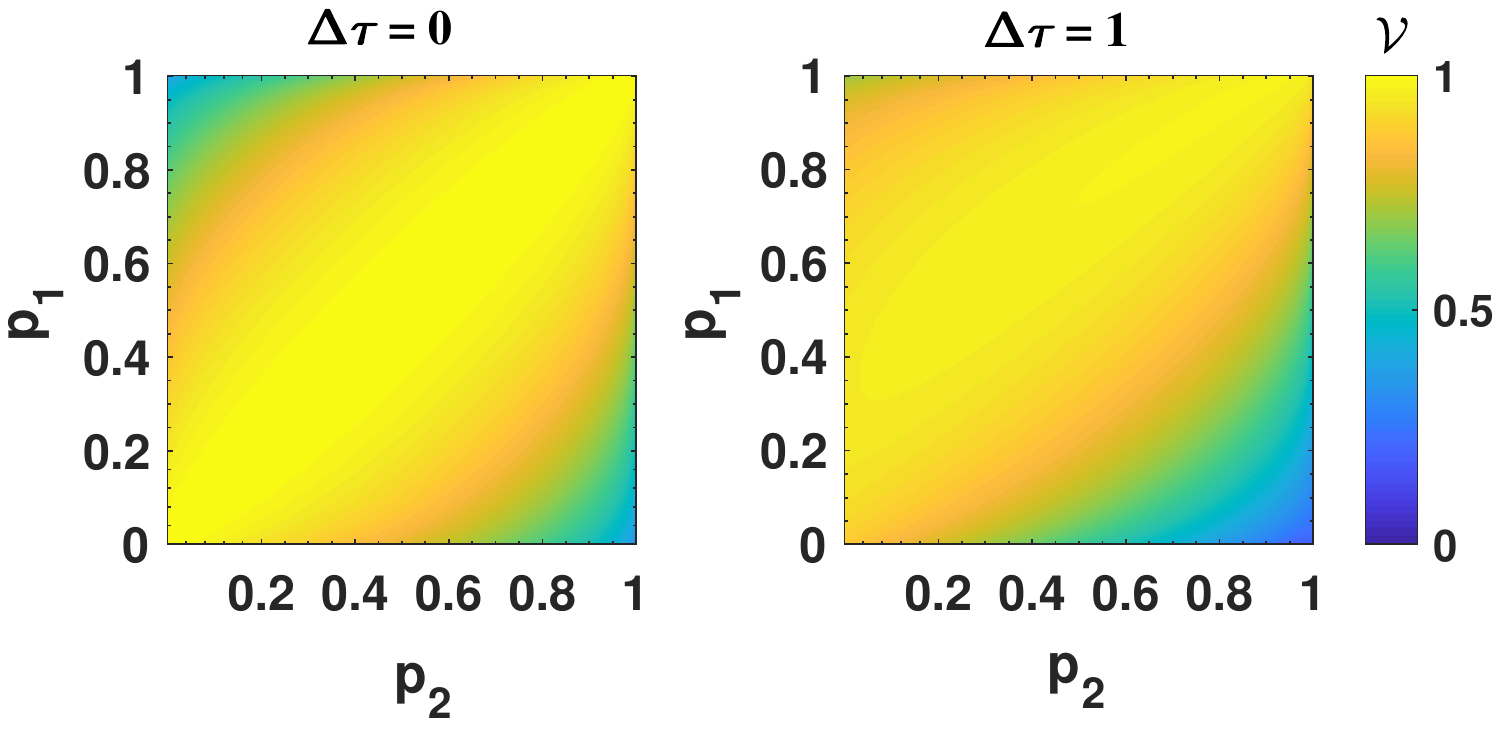}
   \caption{The variation of visibility ($\mathcal{V}$) with $p_1$ and $p_2$, the probabilities of transitions in the arms $\gamma_1$ and $\gamma_2$ such that $\lambda_1 \neq\lambda_2$ for the noise model based on AD channel. The decrease in the visibility is asymmetric and is more with respect to the probability corresponding to the arm with longer transit time. It is assumed that $\tau_1=1$, $\Delta E = 1$ are fixed.}
    \label{fig:AD1}
\end{figure}

We now use this state to obtain the visibility using Eq.~\eqref{eq:diff} with $t=\tau_i$ for propagation along the path $\gamma_i$ of the MZ interferometer. In the two arms of the MZ interferometer, the proper times are different and, for example, are related by the formula for gravitational time dilation for paths following different heights above the earth. As we see from the above equation of the final state, the visibility depends on the difference in proper times $\Delta \tau$ and also on $\Delta E$, $\lambda$.

To further probe the above dependencies, at first, we assume that for the arms 1 and 2, the associated energy scales $\lambda_1 $ and $\lambda_2 $ are the same i.e. $\lambda_1 = \lambda_2$, the resulting visibility is plotted in Fig.~\ref{fig:AD4}. This implies that the environments in the respective arms have a difference in associated probabilities of transition i.e. $p_1$ and $p_2$. The case wherein the energy scale of the environment through the arms is different is shown in Fig.~\ref{fig:AD1}. This case provides for additional distinguishability to the path and hence, the loss of visibility is due to a combination of the effects of time dilation and of noise acting directly on the path.

It is noteworthy that if all other parameters are kept constant and there is no noise ($\lambda=0$), the visibility shows an oscillatory behaviour with the difference in proper time $\Delta \tau$ and the difference in energy level of the two-level system modelled as a clock, i.e.\ $\Delta E$, independently -- see Fig.~\ref{fig:cc} (d). With the inclusion of noise as in AD channel, the situation becomes less intuitive. Generally speaking, the noise tends to disturb the nature of these oscillations disturbing its periodicity, coherence and amplitude. Our interest mainly lies in studying the interplay between the intrinsic dynamics of the clock and the noise, therefore, we analyze the variation of the visibility with $\Delta E$ and $\lambda$  for fixed values of $\Delta \tau$ in Fig.~\ref{fig:AD4}. We see that the interplay between $H_0$ and $H_\textrm{noise}$ leads to the formation of fringes in the visibility pattern.

Next, if we assume that the probabilities through arms 1 and 2 are fixed such that $\lambda_1\neq\lambda_2$, the loss in visibility due to this noise is reflected asymmetrically while changing $p_1$ and $p_2$. In case of an AD channel, we find that there is a marked decrease in visibility for high values of $p_1$ and $p_2$ even when the proper times $\tau_1$ and $\tau_2$ are equal ($\Delta\tau =0$). This decrease is symmetric with respect to $p_1$ and $p_2$. However, when the proper times are different, the decrease is asymmetric  and is more on the side of probability of transition corresponding to the arm with higher proper time. This is shown in Fig.~\ref{fig:AD1}. These effects are due to the added distinguishability offered by the environment due to different $\lambda$'s, reinforcing our understanding of the complementarity between $\mathcal{D}$ and $\mathcal{V}$ (Eq.~\eqref{eq:inequality}). 

\subsection{Phase Damping Channel}

A PD channel is used to model repeated weak interactions of the environment with the target, resulting in loss of coherence between the target qubit's energy eigenstates.  The phase damping channel can be represented using 3 states of the environment in minimum. Therefore, one may assume that the clock interacts with an additional qutrit whose states are given as \{$|1\rangle, |2\rangle, |3\rangle$\}. The unitary time evolution operator governing the dynamics of the extended system in the basis $\{|00\rangle, |01\rangle, |02\rangle, |10\rangle, |11\rangle, |12\rangle\}$ is given as follows:
\begin{align}
U_\textrm{PD}(\tau^*) = \begin{bmatrix}
            \sqrt{1-p}&-\sqrt{p}&0&0&0&0\\
            \sqrt{p}&\sqrt{1-p}&0&0&0&0\\
            0&0&1&0&0&0\\
            0&0&0&\sqrt{1-p}&0&-\sqrt{p}\\
            0&0&0&0&1&0\\
            0&0&0&\sqrt{p}&0&\sqrt{1-p}
\end{bmatrix}.
\end{align}
The eigenvalues (with the substitution $p=\sin^2\theta$) are: $\{1,1,e^{-i\theta}, e^{-i\theta},e^{i\theta},e^{i\theta}\}$ and the corresponding eigenvectors are:
\begin{alignat}{3}
&|v_1'\rangle=|11\rangle ,~&&|v_2'\rangle=|02\rangle,~
& &|v_3'\rangle=-i|10\rangle+|12\rangle ,\nonumber\\
&|v_4'\rangle=-i|00\rangle+|01\rangle,~
&&|v_5'\rangle=i|10\rangle +|12\rangle ,~& &|v_6'\rangle=i|00\rangle+|01\rangle .
\end{alignat}
The eigenvectors of the Hamiltonian governing the dynamics of the clock system are the same as above and the eigenvalues are:  $\bigg\{0,0,\frac{-\theta}{\tau^*}, \frac{-\theta}{\tau^*}, \frac{\theta}{\tau^*},\frac{\theta}{\tau^*}\bigg\}$.
Hence, the Hamiltonian can be expressed as follows:
\begin{align}
H_\textrm{PD} =\frac{-\theta}{\tau^*}\bigg[|v_3'\rangle\langle v_3'|+|v_4'\rangle\langle v_4'|-|v_5'\rangle\langle v_5'|
-|v_6'\rangle\langle v_6'|\bigg]\equiv 2i\lambda \big[-|12\rangle\langle 10|+|10\rangle\langle12| -|01\rangle\langle00|+|00\rangle\langle01|\big],  
\label{eq:HPD}
\end{align}
where $\lambda=\theta/\tau^*$. Adding the system's internal Hamiltonian, Eq.~\eqref{eq:H0}, we obtain for the combined system-environment Hamiltonian as follows:
\begin{align}
  H_\textrm{int}=H_{0}+H_\textrm{PD} = 
\begin{bmatrix}
         E_0&2i\lambda&0&0&0&0\\
         -2i\lambda&E_0&0&0&0&0\\
         0&0&E_0&0&0&0\\
         0&0&0&E_1&0&2i\lambda\\
         0&0&0&0&E_1&0\\
         0&0&0&-2i\lambda&0&E_1
\end{bmatrix}.
\end{align}
The eigenvalues of the above Hamiltonian are: $\{E_0,E_1,E_0-2\lambda,E_1-2\lambda,E_0+2\lambda,E_1+2\lambda\}$. The corresponding eigenvectors are labelled as $\{|v_1\rangle,|v_2\rangle,|v_3\rangle,|v_4\rangle,|v_5\rangle,|v_6\rangle\}$. Using the above, the initial state of the clock and environment can be expressed in the eigenvector basis as follows:
\begin{equation}
|\psi\rangle_\textrm{CE}(t=0) = \frac{i}{2\sqrt{2}}\big[|v_3\rangle+|v_4\rangle-|v_5\rangle-|v_6\rangle\big].
\end{equation}
The time evolution of the initial state to a general time $t$ gives:
\begin{align}
|\psi\rangle_\textrm{CE}(t) = 
         \frac{i e^{-i~(E_0-2\lambda)~t}}{2\sqrt2}|v_3\rangle+\frac{i e^{-i~(E_1-2\lambda)~t}}{2\sqrt2}|v_4\rangle -\frac{i e^{-i~(E_0+2\lambda)~t}}{2\sqrt2}|v_5\rangle-\frac{i e^{-i~(E_1+2\lambda)~t}}{2\sqrt2}|v_6\rangle . 
\end{align}
\begin{figure}
    \centering
    \includegraphics[width=0.5\columnwidth]{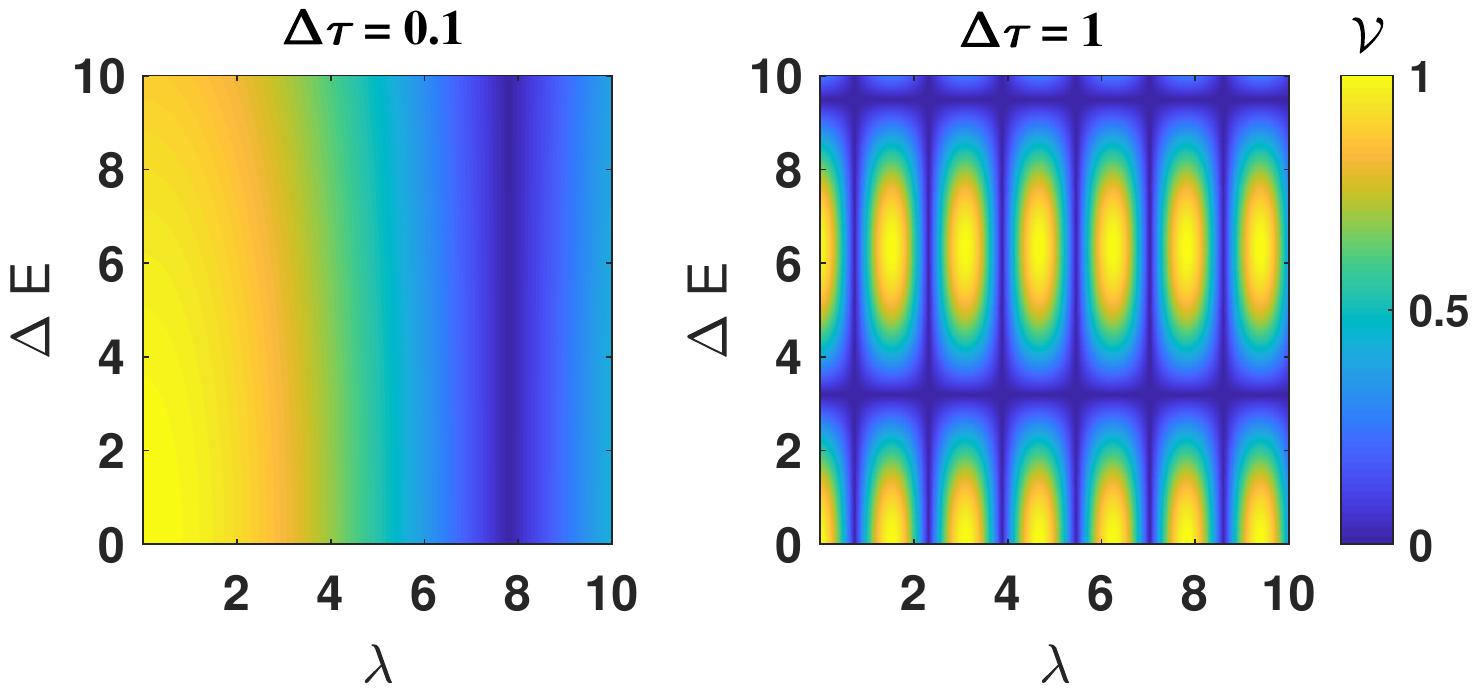}
   \caption{The variation of $\mathcal{V}$ with $\Delta E$ and $\lambda$ for various values of $\Delta \tau$ for the noise model based on PD channel. The pattern of $\mathcal{V}$ changes in an oscillatory manner with $\Delta \tau$, $\Delta E$ and $\lambda$ independently. The introduction of phase damping channel leads to the formation of fringes in the visibility pattern. In this case, $\lambda_1=\lambda_2$.}
    \label{fig:DP4}
    \centering
    \includegraphics[width=0.5\columnwidth]{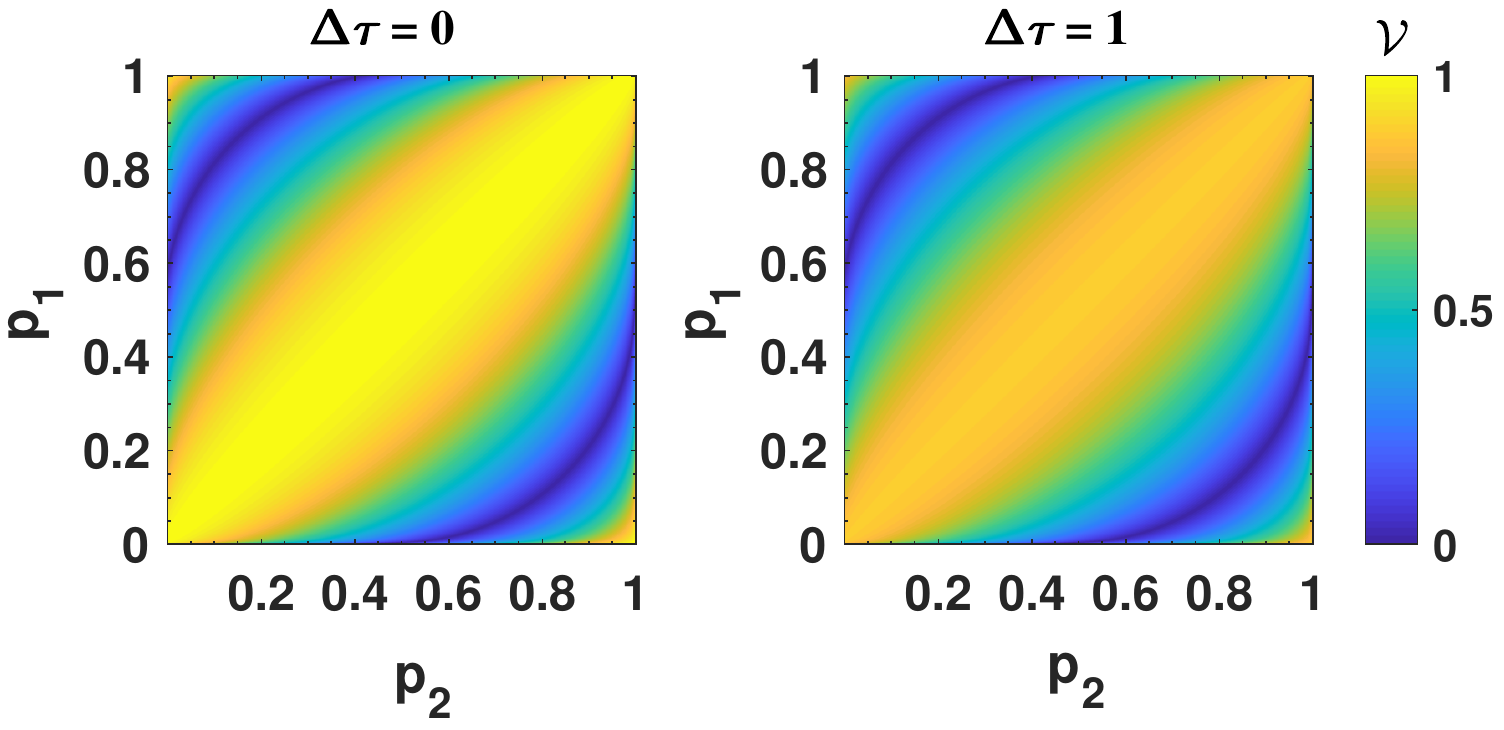}
   \caption{Variation of $\mathcal{V}$ with $p_1$ and $p_2$ in the case when $\lambda_1\neq\lambda_2$ for the noise model based on PD channel. In this case, increased $\Delta \tau$ leads to uniform decrease of $\mathcal{V}$ with respect to the probabilities in both the arms. The behaviour is, therefore, symmetric in contrast to AD channel (Fig.~\ref{fig:AD1}) which shows an asymmetric behaviour. We have assumed that $\tau_1=1$ and $\Delta E=1$ are fixed.}
    \label{fig:DP1}
\end{figure}
Now, we use the above state to calculate the visibility $\mathcal{V}$ using Eq.~\eqref{eq:diff}.
It is seen in Fig.~\ref{fig:DP4} that the interplay between the noise represented by $\lambda$ and the system's energy gap $\Delta E$ leads to the formation of fringes in the visibility pattern. This has been considered under the assumption that $\lambda_1=\lambda_2$. In Fig.~\ref{fig:DP1}, we  consider $\lambda_1\neq\lambda_2$. As opposed to the effect of AD channel which shows distinction in the visibility with respect to a longer proper time as seen in Fig.~\ref{fig:AD1}, the effect of PD channel is such that the change in visibility pattern due to changing the difference in proper time ($\Delta \tau$) is symmetric.
\subsection{Depolarizing Channel}

A DP channel introduces a slew of errors on the target qubit by a combination of Pauli operators: $\sigma_x$, $\sigma_y$ and $\sigma_z$. The underlying transitions that model noise due to a DP channel are presented in a generic form in Eq.~\eqref{DP}. Depending on whether a transition occurs and which transition occurs, the environment assumes either of the 4 states : $\{|0\rangle,|1\rangle,|2\rangle,|3\rangle \}$.

\begin{align}
  U_\textrm{DP} \big(|\psi\rangle_\textrm{C} \otimes |0\rangle_\textrm{E}\big)= \sqrt{1-p} |\psi\rangle_\textrm{C}\otimes |0\rangle_\textrm{E}+\sqrt{\frac{p}{3}}\bigg[\sigma_x|\psi\rangle_\textrm{C}\otimes |1\rangle_\textrm{E} +\sigma_y|\psi\rangle_\textrm{C}\otimes|2\rangle_\textrm{E}+\sigma_z|\psi\rangle_\textrm{C}\otimes |3\rangle_\textrm{E}\bigg].
  \label{DP}
\end{align}

In the basis $\{|00\rangle, |01\rangle, |02\rangle, |03\rangle, |10\rangle, |11\rangle, |12\rangle, |13\rangle \}$, where the left and right label in the ket represent the clock and the environment respectively, the unitary time evolution operator governing the dynamics of the extended system (clock + environment) is of the following form:
\begin{align}
U_\textrm{DP} (\tau^*)=
\begin{bmatrix}
 \sqrt{1 - p}&0& 0& \frac{i\sqrt{p}}{\sqrt{3}}& 0& \frac{i\sqrt{p}}{\sqrt{3}}& \frac{\sqrt{p}}{\sqrt{3}}& 0\\
 0& \sqrt{1 - p}& \frac{i\sqrt{p}}{\sqrt{3}}& 0& \frac{i\sqrt{p}}{\sqrt{3}}& 0&   0& \frac{\sqrt{p}}{\sqrt{3}}\\
 0& \frac{i\sqrt{p}}{\sqrt{3}}& \sqrt{1 - p}& 0& \frac{\sqrt{p}}{\sqrt{3}}& 0&   0& \frac{i\sqrt{p}}{\sqrt{3}}\\
 \frac{i\sqrt{p}}{\sqrt{3}}& 0& 0& \sqrt{1 - p}& 0& \frac{\sqrt{p}}{\sqrt{3}}& \frac{i\sqrt{p}}{\sqrt{3}}& 0\\
 0& \frac{i\sqrt{p}}{\sqrt{3}}& \frac{-\sqrt{p}}{\sqrt{3}}& 0& \sqrt{1 - p}& 0&  0& \frac{-i\sqrt{p}}{\sqrt{3}}\\
 \frac{i\sqrt{p}}{\sqrt{3}}& 0& 0& \frac{-\sqrt{p}}{\sqrt{3}}& 0& \sqrt{ 1 - p}& \frac{-i\sqrt{p}}{\sqrt{3}}& 0\\
 \frac{-\sqrt{p}}{\sqrt{3}}& 0& 0& \frac{i\sqrt{p}}{\sqrt{3}}&  0& \frac{-i\sqrt{p}}{\sqrt{3}}& \sqrt{1 - p}& 0\\
 0& \frac{-\sqrt{p}}{\sqrt{3}}& \frac{i\sqrt{p}}{\sqrt{3}}&  0& \frac{-i\sqrt{p}}{\sqrt{3}}& 0& 0& \sqrt{1 - p}
\end{bmatrix}.
\label{UDP}
\end{align}

The eigenvalues (with substitution $p=\sin^2 \theta$) are:
\{ $e^{-i\theta}$, $e^{-i\theta}$, $e^{-i\theta}$, $e^{-i\theta}$, $e^{i\theta}$, $e^{i\theta}$, $e^{i\theta}$, $e^{i\theta}$\}. 

The corresponding eigenvectors in the same basis are as follows:
\begin{alignat}{2}
|v_1\rangle&=|v_3\rangle= i\sqrt{3}|01\rangle-i|02\rangle-i|10\rangle+|13\rangle,~&&|v_2\rangle =|v_4\rangle= i\sqrt{3}|00\rangle-i|03\rangle-i|11\rangle+|12\rangle, \nonumber\\
|v_5\rangle &=|v_7\rangle= -i\sqrt{3}|01\rangle-i|02\rangle-i|10\rangle+|13\rangle,~
&&|v_6\rangle =|v_8\rangle= -i\sqrt{3}|00\rangle-i|03\rangle-i|11\rangle+|12\rangle . \nonumber\\
\end{alignat}
Therefore, the Hamiltonian for the noise model based on DP channel is written as follows:
\begin{align}
H_\textrm{DP} &=\frac{\theta}{\tau^*}\big[|v_1\rangle\langle v_1|+|v_2\rangle\langle v_2| +|v_3\rangle\langle v_3|+|v_4\rangle\langle v_4| -|v_5\rangle\langle v_5|-|v_6\rangle\langle v_6| -|v_7\rangle\langle v_7|-|v_8\rangle\langle v_8|\big]\nonumber\\
&\equiv 4\lambda \big[i\sqrt{3}|12\rangle\langle 00|+\sqrt{3}|11\rangle\langle00|+\sqrt{3}|03\rangle\langle00|
+i\sqrt{3}|13\rangle\langle01|+\sqrt{3}|10\rangle\langle01|+\sqrt{3}|02\rangle\langle01|\nonumber\\
&+\sqrt{3}|00\rangle\langle11|+\sqrt{3}|00\rangle\langle03|-i\sqrt{3}|00\rangle\langle12| 
-i\sqrt{3}|01\rangle\langle13|+\sqrt{3}|01\rangle\langle10|+\sqrt{3}|01\rangle\langle02|\big],
\label{eq:DPCE}
\end{align}
where $\lambda= \theta/\tau^*$ which can again be interpreted as the energy scale of the noise Hamiltonian. The complete Hamiltonian can be written by adding $H_0$ (Eq.~\eqref{eq:H0}) to $H_\textrm{DP}$:
\begin{equation}
  H_\textrm{int}=H_{0}+H_\textrm{DP} = 
\begin{bmatrix}
         E_0&0&0&4\sqrt{3}\lambda&0&4\sqrt{3}\lambda&-4i\sqrt{3}\lambda&0\\
         0&E_0&4\sqrt{3}\lambda&0&4\sqrt{3}\lambda&0&0&-4i\sqrt{3}\lambda\\
         0&4\sqrt{3}\lambda&E_0&0&0&0&0&0\\
         4\sqrt{3}\lambda&0&0&E_0&0&0&0&0\\
         0&4\sqrt{3}\lambda&0&0&E_1&0&0&0\\
         4\sqrt{3}\lambda&0&0&0&0&E_1&0&0\\
         4i\sqrt{3}\lambda&0&0&0&0&0&E_1&0\\
         0&4i\sqrt{3}\lambda&0&0&0&0&0&E_1\\
\end{bmatrix}.
\label{HDPCE}
\end{equation}

\begin{figure}
    \centering
    \includegraphics[width=0.5\columnwidth]{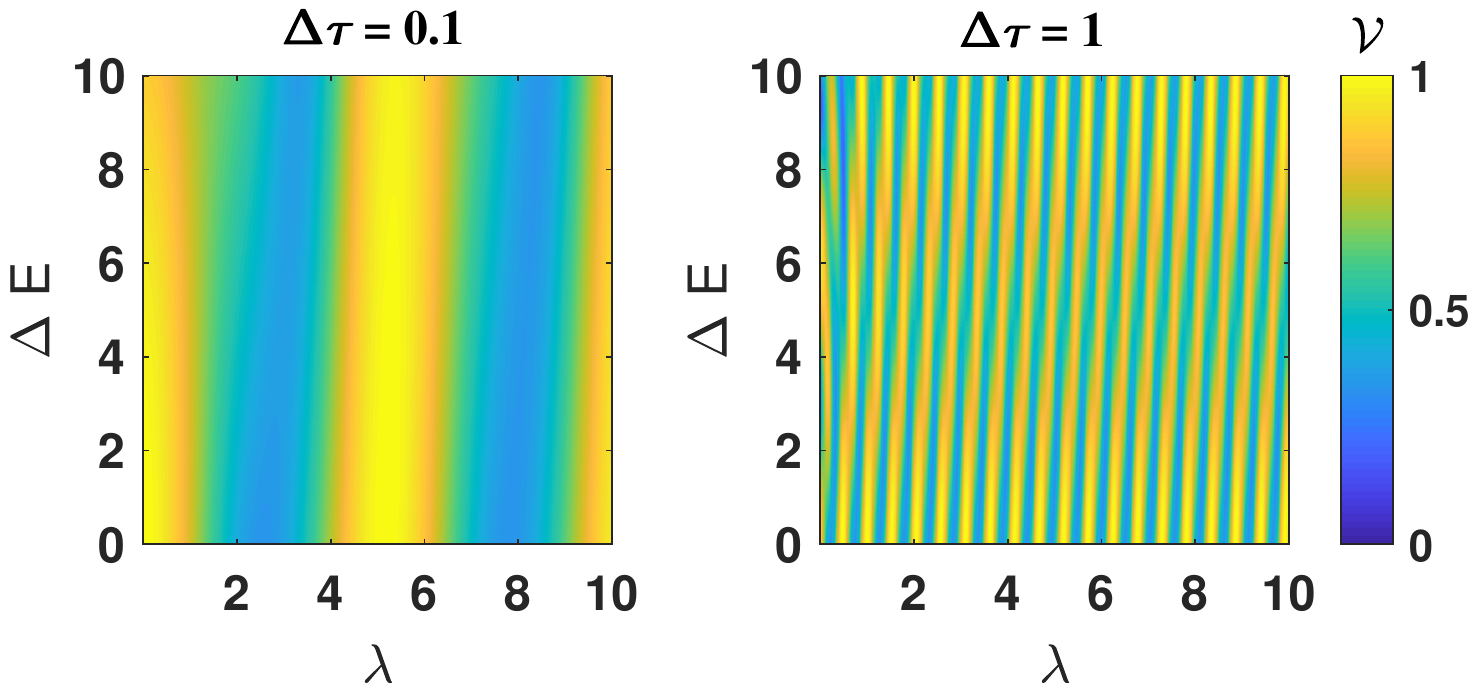}
   \caption{The variation of $\mathcal{V}$ with of $\lambda$ and $\Delta E$ for the noise model based on DP channel. The interplay between these leads to the formation of fringes in the visibility pattern. In this case, $\lambda_1=\lambda_2$.} 
    \label{fig:PD4}
    \centering
    \includegraphics[width=0.5\columnwidth]{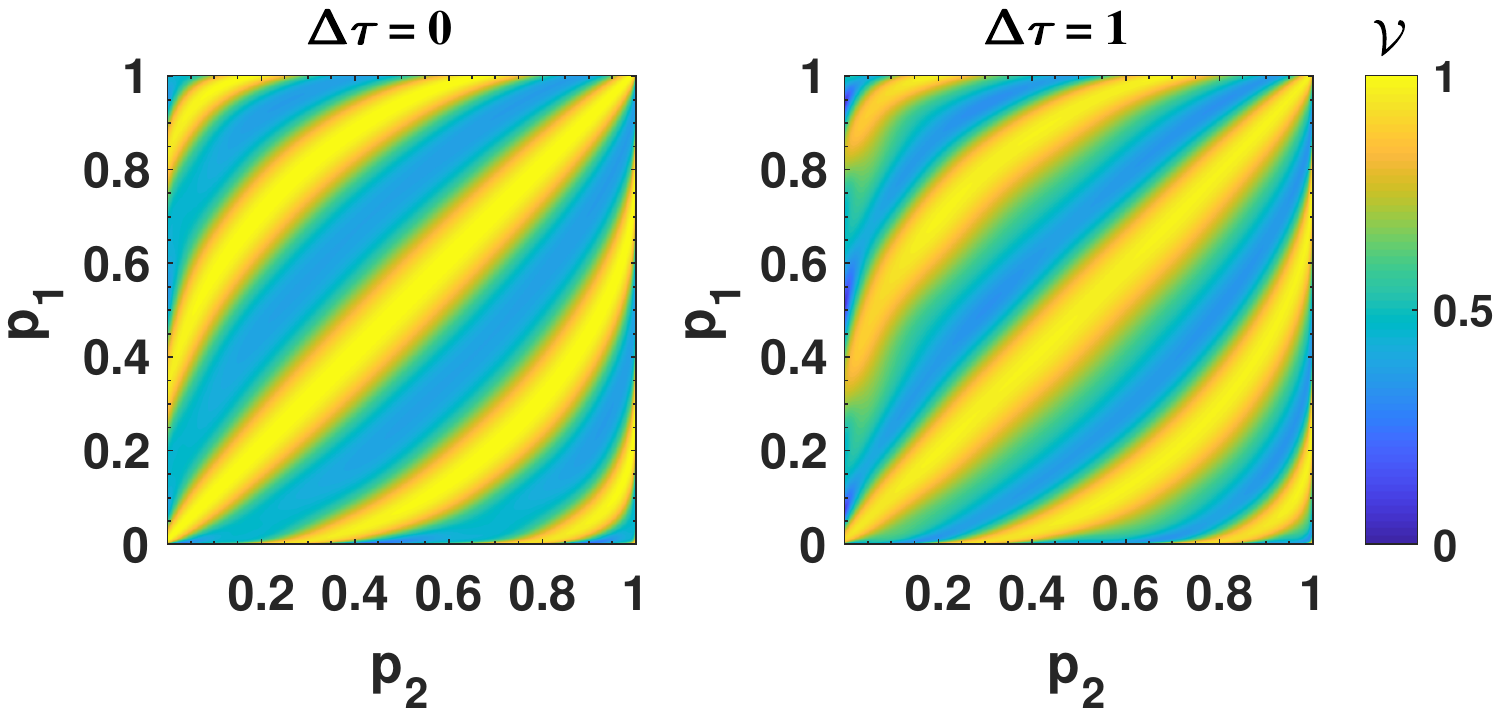}
   \caption{$\mathcal{V}$ as a function of probabilities $p_1$ and $p_2$ for the noise model based on DP channel. There is a marked decrease of $\mathcal{V}$ in high the visibility regions as $\Delta \tau$ is increased. For this case, $\lambda_1\neq\lambda_2$ has been assumed and $\tau_1=1$ and $\Delta E=1$ are fixed.}
    \label{fig:PD1}
\end{figure}

Owing to the involved nature of analytical calculations required for calculating the time evolved state of the clock, we have undertaken a numerical route for calculating the visibility in the case of the DP channel. Taking the matrix exponential of the above Hamiltonian,  $e^{-iH_\textrm{int}\tau_{i}}$ and multiplying it with the initial vector $\big[\frac{1}{\sqrt{2}},0,0,0,\frac{1}{\sqrt{2}},0,0,0\big]$, given in Eq.~\eqref{eq:init} gives the time evolved state from which we calculate the visibility using Eq. \eqref{eq:visi}. We consider the case with $\lambda_1=\lambda_2$ in Fig.~\ref{fig:PD4} for different $\Delta \tau$. The interplay between $\lambda$ and $\Delta E$ is captured. The fringes are formed due to the oscillatory nature of the visibility with $\Delta E$, $\Delta \tau$ and $\lambda$ independently. In Fig.~\ref{fig:PD1}, we consider the case when $\lambda_1\neq\lambda_2$, which results in a symmetric visibility pattern, similarly to the PD channel in Fig.~\ref{fig:DP1}.

\begin{figure}
   \centering
    \includegraphics[scale=0.5,width=0.5\columnwidth]{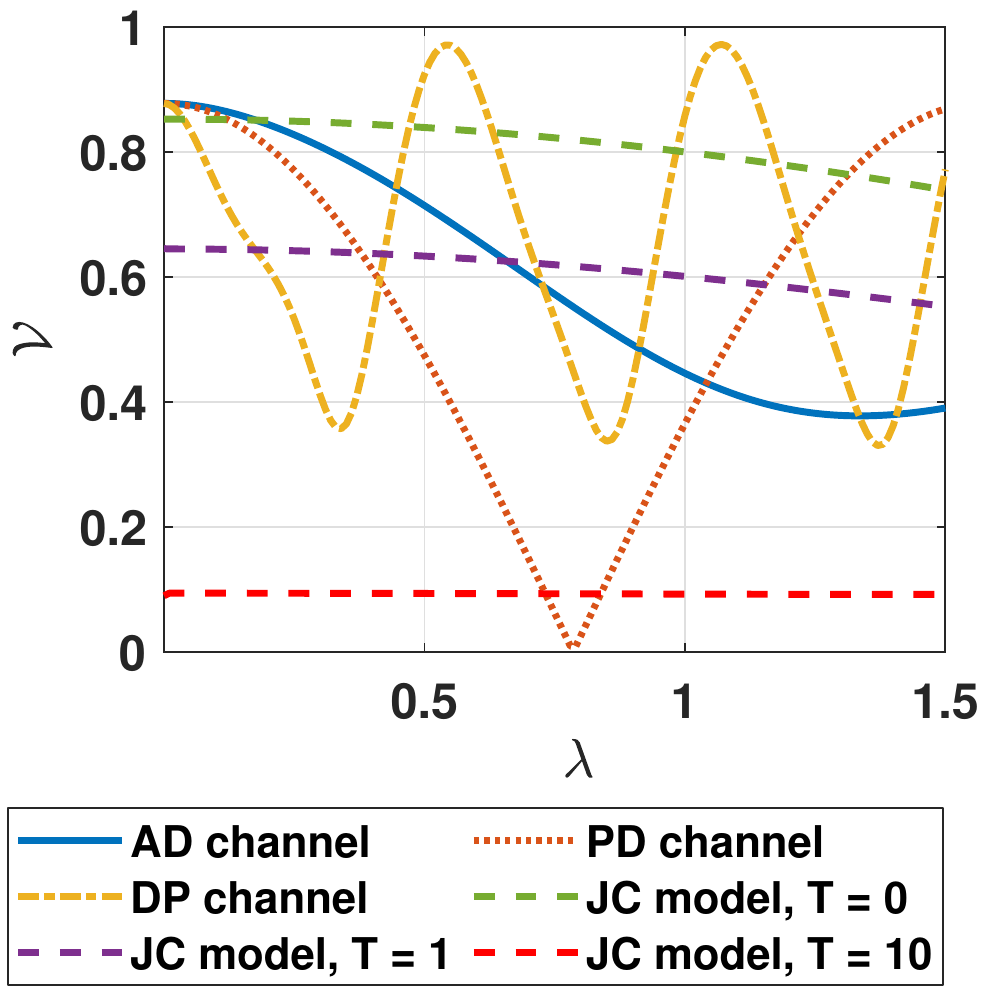}
   \caption{Comparison of $\mathcal{V}$ vs. $\lambda$ for the JC model, noise models based on AD, PD and DP  channels. Increasing $\lambda$ signifies higher energy scale of interactions in the noise Hamiltonian. Here, $\Delta E = 1$ and $\omega = 1.1$ for the JC model. $\Delta E = 1$ for all others and $\Delta \tau = 1$ is fixed in all the cases. In the absence of noise, one would expect a constant line at $\mathcal{V}=0.8775$ with an environment without self interaction as can be seen for all noise models except the JC model for which $\omega \neq 0$. $\lambda_1=\lambda_2$ ($\equiv \lambda$) has been assumed in all the cases.}
    \label{fig:C1}
\end{figure}
\begin{figure*}[h!]
    \centering
    \includegraphics[width=\columnwidth]{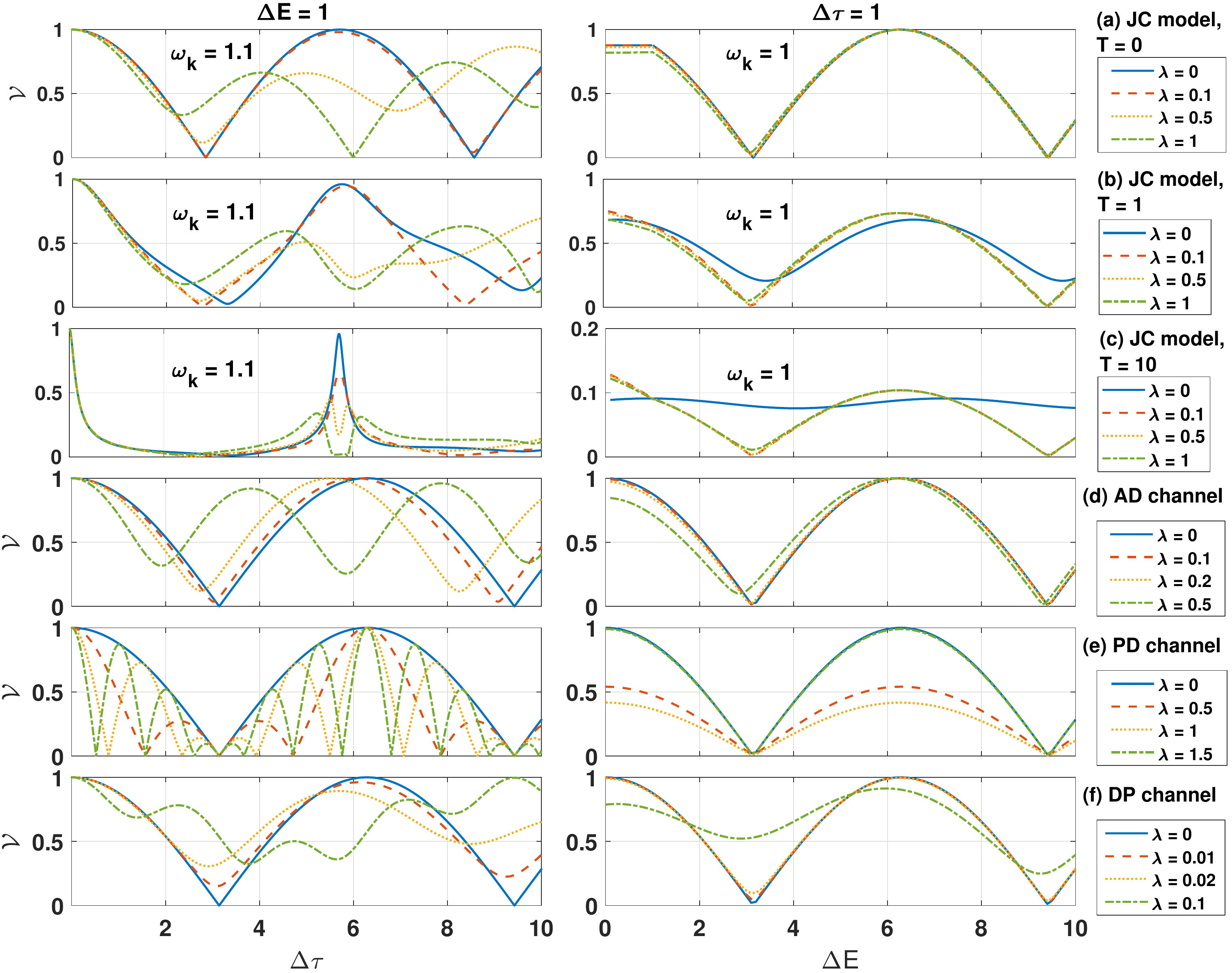}
   \caption{Comparison of $\mathcal{V}$ vs. the difference in proper time ($\Delta \tau$) for (a), (b), (c) JC Model and other noise models based on (d) AD, (e) PD and (f) DP channels separately under various scale of energies of the noise Hamiltonian ($\lambda$). $\Delta E =1$ has been fixed for all values of $\lambda$, for the plots shown in the left column. Comparison of $\mathcal{V}$ vs. $\Delta E$ for various values of $\lambda$, with $\Delta \tau =1$ is shown in the right column. There is a marked difference in the nature of changes in the visibility depending on the type of noise model and $\lambda$. Different $\lambda$ scales have been selected to be shown due to the differing susceptibilities of $\mathcal{V}$ to $\lambda$. $\lambda_1=\lambda_2$ ($\equiv \lambda$) has been assumed in all cases.}
    \label{fig:cc}
\end{figure*}

\section{Comparison between channels}

We shall now compare the visibility obtained in the different noise models based on the JC model, AD, PD and DP channels while keeping two of the parameters among $\Delta E$, $\lambda$ and $\Delta \tau$ constant. At first in Fig.~\ref{fig:C1}, we look at the case wherein $\Delta \tau =1$, $\Delta E=1$ have been fixed and $\lambda$ takes the values from $0$ to $1.5$. Qualitatively, there is a marked difference between the behaviour of the visibility among the different cases as $\lambda$ is increased. In the low noise regime, i.e., for small $\lambda$, we inferred that the visibility should decrease due to the dominance of the effect of a larger Hilbert space over the effect of noise. This is indeed observed in Fig.~\ref{fig:C1}. Moreover, in the minimalist scheme for the modelling of environment that we have considered, the noise model based on DP channel -- which requires four states of the environment for its description shows a steeper decrease in visibility in this regime than that based on the PD channel which requires a three level environment. Similarly, for the model based on an AD channel which requires two levels of the environment in the minimal representation, we see the lowest decrease in the visibility, again adhering to our argument. Therefore, we observe a trend $\mathcal{V}_\textrm{DP}<\mathcal{V}_\textrm{PD}<\mathcal{V}_\textrm{AD}$ in the low noise regime in Fig.~\ref{fig:C1} as expected. In the general regime, the noise model based on DP channel shows distorted fast oscillations in the visibility whereas the one based on a PD channel shows large oscillations with the visibility even falling to $0$. In contrast, the noise model based on AD channel shows slower and lesser amplitude oscillations in the visibility. In the case of a PD channel under appropriate conditions, it leads to zero visibility which is also evident from the blue regions in the fringe pattern of the visibility as seen in Fig.~\ref{fig:DP4}. Interestingly, for the DP channel based noise model, we see that for certain $\lambda$'s the visibility is higher than that in the case without the environment.

The case of JC model with various temperatures is qualitatively different than the others as it also includes an intrinsic Hamiltonian of the environment field ($H_\textrm{env}$) which governs its time evolution (if $\omega \neq 0$, as considered). Because of this fact, at $\lambda=0$ it shows a lower visibility than the channel based noise models, since even without the noise Hamiltonian the clock and the environment evolve governed by their respective time evolution operators -- $H_0$, and $H_\textrm{env}$ respectively. In general, it is found that for low $\lambda$, a lower visibility is obtained at higher temperatures.

Next, we ascertain the effect of noise by analyzing the variation of the visibility with the difference in proper time ($\Delta \tau$) in Fig.~\ref{fig:cc}, left column, and with the difference in energy levels of the clock ($\Delta E$) while keeping the other quantities fixed in Fig.~\ref{fig:cc}, right column. Without the effect of environment, the visibility shows a periodic behaviour with both $\Delta E$ and $\Delta \tau$ with a period of $\pi$. While this translates to $\lambda = 0$ for channel based noise models, the JC model would require $\omega = 0$ in addition to $\lambda=0$ for a noise free evolution of the system (clock+environment). This explains why Fig.~\ref{fig:cc} (a), (b), (c) have a different behaviour for $\lambda = 0$. It is noteworthy that we have used different scales of $\lambda$ for different noise models depending on the susceptibility of the visibility to $\lambda$. The DP channel based noise model shows the most change in the visibility even for small values of $\lambda$ as is evident from Fig.~\ref{fig:cc} (f). For an AD channel based noise model, as seen from Fig.~\ref{fig:cc} (d), this scale is intermediate with $\lambda\in[0,0.5]$. For this case, there are large deviations in the case of $\lambda=0.5$ for fixed $\Delta E$ whereas the deviations are small for fixed $\Delta \tau$. For a PD channel based noise model, as seen from Fig.~\ref{fig:cc} (e), the introduction of environment leads to change in the periodicity and amplitude of the oscillations of visibility with respect to $\Delta \tau$ and $\Delta E$ respectively. Importantly, we again observe that noise models based on AD and DP channels result in the visibility higher than that in the case without the environment (solid blue lines in all the plots) for certain values of $\lambda$. A similar phenomenon is also seen for the JC model though $\lambda = 0$ there doesn't correspond to the case without environment, as outlined before.
\end{document}